\documentclass{jfm}

\usepackage{graphicx}
\usepackage{natbib}
\usepackage{color}
\usepackage{amsmath}
\usepackage{epsfig}
\usepackage{lineno}
\usepackage{subfigure}
\usepackage{caption}
\usepackage[normalem]{ulem}
\usepackage{gensymb}

\DeclareFontFamily{OT1}{pzc}{}
\DeclareFontShape{OT1}{pzc}{m}{it}{<-> s * [1.10] pzcmi7t}{}
\DeclareMathAlphabet{\mathpzc}{OT1}{pzc}{m}{it}

\ifCUPmtlplainloaded \else
  \checkfont{eurm10}
  \iffontfound
    \IfFileExists{upmath.sty}
      {\typeout{^^JFound AMS Euler Roman fonts on the system,
                   using the 'upmath' package.^^J}%
       \usepackage{upmath}}
      {\typeout{^^JFound AMS Euler Roman fonts on the system, but you
                   dont seem to have the}%
       \typeout{'upmath' package installed. JFM.cls can take advantage
                 of these fonts,^^Jif you use 'upmath' package.^^J}%
      }
  \else
  \fi
\fi

\ifCUPmtlplainloaded \else
  \checkfont{msam10}
  \iffontfound
    \IfFileExists{amssymb.sty}
      {\typeout{^^JFound AMS Symbol fonts on the system, using the
                'amssymb' package.^^J}%
       \usepackage{amssymb}%
         \let\leq=\leqslant
         \let\geq=\geqslant
      }{}
  \fi
\fi

\ifCUPmtlplainloaded \else
  \IfFileExists{amsbsy.sty}
    {\typeout{^^JFound the 'amsbsy' package on the system, using it.^^J}%
     \usepackage{amsbsy}}
    {\providecommand\boldsymbol[1]{\mbox{\boldmath $##1$}}}
\fi

\newcommand\Pe{\mbox{\textit{Pe}}}  

\newsavebox{\astrutbox}
\sbox{\astrutbox}{\rule[-5pt]{0pt}{20pt}}

\def\XXint#1#2#3{{\setbox0=\hbox{$#1{#2#3}{\int}$}
     \vcenter{\hbox{$#2#3$}}\kern-.5\wd0}}

\newcommand\etal{\mbox{\textit{et al.}}}

\newcommand\Ca{\mbox{\textit{Ca}}}

\newcommand{\ve}{\varepsilon}
\newcommand{\skn}{\mathring}
\newcommand{\tit}{\textit}
\newcommand{\tbf}{\textbf}
\newcommand{\mbf}{\boldsymbol}

\title[The nascent coffee ring with an arbitrary droplet contact set]{The nascent coffee ring with arbitrary droplet contact set: an asymptotic analysis}

\author[M. R. Moore \etal]%
{M.\ns R.\ns M\ls O\ls O\ls R\ls E$^{1,2}$, \ns 
D.\ns V\ls E\ls L\ls L\ls A$^2$ \ns
\and
J.\ns M.\ns O\ls L\ls I\ls V\ls E\ls R$^2$}

\affiliation{$^1$Department of Physics \& Mathematics, University of Hull, Cottingham Road, Kingston-upon-Hull, HU6 7RX
\\
$^2$Mathematical Institute, University of Oxford, Andrew Wiles Building, Radcliffe Observatory Quarter, Woodstock Road, Oxford, OX2 6GG}

\pubyear{}
\volume{}
\pagerange{}
\date{?; revised ?; accepted ?. - To be entered by editorial office}

\begin{document}

\maketitle



\begin{abstract}
We consider the effect of droplet geometry on the early-stages of coffee ring formation during the evaporation of a thin droplet with an arbitrary simple, smooth, pinned contact line. We perform a systematic matched asymptotic analysis of the small-capillary number, large-solutal P\'{e}clet number limit for two different evaporative models: a kinetic model, in which the evaporative flux is effectively constant across the droplet, and a diffusive model, in which the evaporative flux is singular at the contact line. For both evaporative models, solute is transported to the contact line by a capillary flow in the bulk of the droplet while, local to the contact line, solute diffusion counters advection. The resulting interplay leads to the formation of the nascent coffee ring profile. By exploiting a coordinate system embedded in the contact line, we solve explicitly the local leading-order problem, deriving a similarity profile (in the form of a gamma distribution) that describes the coffee ring profile in its early stages. Notably, for an arbitrary contact line geometry, the ring characteristics change due to the concomitant asymmetry in the shape of the droplet free surface, the evaporative flux (in the diffusive evaporative regime) and the mass flux into the contact line. We utilize the asymptotic model to determine the effects of contact line geometry on the growth of the coffee ring for a droplet with an elliptical contact set. Our results offer mechanistic insight into the effect of contact-line curvature on the development of the coffee-ring from deposition up to jamming of the solute; moreover our model predicts when finite concentration effects become relevant.
\end{abstract}

\section{Introduction} \label{sec:Introduction}

The `coffee-ring effect' takes its name from the ringlike deposits left behind on a surface after a spilled droplet of coffee has evaporated into the surrounding air. As uncovered by the seminal work of \citet{Deegan1997, Deegan2000} for an axisymmetric droplet, surface inhomogeneities tend to pin the circular contact line of the droplet so that, to replenish fluid lost during evaporation, an outward radial capillary flow develops inside the droplet; this flow carries solute with it, leading to a build up at the droplet edge. Eventually the solute becomes sufficiently concentrated that particle concentration effects such as increased local suspension viscosity \cite[][]{Kaplan2015} and jamming \cite[][]{Popov2005} become relevant, forming the coffee ring that remains once all liquid has evaporated.

This behaviour is not limited to just coffee, but is ubiquitous whenever a liquid droplet containing a solute is left to evaporate. It is vital to understand how to control the effect in many different biological, industrial and engineering settings \cite[][]{Anyfantakis2015}. There are a number of different mechanisms that can be employed to control the shape of the deposit \cite[][]{Mampallil2018}. A common approach to inhibit coffee ring formation is to interfere with the pinning of the contact line by, for example, employing superhydrophobic substrates \cite[see, for example,][]{Cui2012, Dicuangco2014}, using electrowetting to encourage contact line slippage \cite[for example,][]{Li2008} or utilizing an oil-coated substrate \cite[for example,][]{Li2020}. Alternatively, one can encourage physical effects that counter the outward capillary flow, such as Marangoni effects \cite[see, for example,][]{Hu2006, Ristenpart2007,Li2015} or exploiting electrostatic or electroosmotic controls \cite[for example,][]{Wray2014,Kim2006}. One can also introduce other liquids to the drop: the evaporation of binary droplets is a more complex phenomenon, exhibiting several different flow stages depending on, for example, the relative volatility of the liquid components, which can lead to an interesting variety of deposit patterns \cite[see, for example,][and the references therein]{Kim2016, Zhong2016, Li2018,Pahlavan2021}.

In this paper, we consider the role of droplet \textit{geometry} on the coffee-ring effect. The geometry of a sessile drop can readily be controlled in a laboratory setting, which makes it a valuable tool for potential control of the deposition pattern. For example, a droplet on a sloped surface will be perturbed away from a spherical cap profile by gravity, which leads to a change in the angular dependence of the evaporative flux \cite[][]{Timm2019, Tredenick2021}. The droplet contact set can also be manipulated by machining or treating the substrate in such a way that pinning at particular points is promoted, altering the shape of the deposit \cite[][]{He2017,Saenz2017,Kubyshkina2020}. 

Once asymmetry is introduced, there is no longer uniformity in the coffee-ring profile. \citet{Deegan1997} noted that coffee stains tended to be darker near more curved regions of the contact line of a drying droplet, stating that the (diffusive) evaporative flux is larger in these regions, which in turn drives a stronger capillary flow. \citet{Saenz2017} performed a range of experiments and simulations of different shaped contact sets suggesting that, in addition to the contact line curvature, the mean curvature of the droplet free surface plays an integral role in the evaporation rate. 

While an increased evaporation rate near highly-curved parts of the contact line certainly contributes to an enhanced transport of solute mass into these regions, it is not a necessary requirement. \citet{FreedBrown2015} conducted a numerical investigation of the mass flux of solute into the contact line for a wide range of different droplet profiles evaporating under a uniform evaporative flux. Even without spatially-varying evaporative flux, the liquid velocity is still enhanced towards the highly-curved parts of the boundary and so there is a greater mass flux of solute to this part of the boundary, again acting to strengthen the coffee-ring effect.

In the present work, we extend our recent analysis \cite[][]{Moore2021} to gain an understanding of the early stages of coffee-ring formation for arbitrary contact line geometries. In the model of \citet{Deegan1997, Deegan2000}, the solute is sufficiently dilute in the fluid that it is advected radially outwards by the capillary flow, with all of the solute concentrated in an infinitesimally-small ring at the contact line once the liquid has completely evaporated. Naturally, real coffee rings have finite dimensions and the Deegan model can be adapted to incorporate the effects of finite solute concentration \cite[see, for example,][]{Popov2005,Kaplan2015}. However, an important aspect of the dilute problem that is missing in the Deegan model is the effect of solute diffusion, which resists the development of large spatial gradients induced by particle advection close to the contact line. For a thin, axisymmetric droplet, \citet{Moore2021} performed an asymptotic analysis in the physically-relevant limit of small-capillary number and large-solutal P\'{e}clet number to show that, by including the effects of diffusion, the characteristic narrow, peaked coffee-ring develops even in the early stages of evaporation, for which the solute remains dilute. The local form of the coffee-ring was shown to collapse onto a universal gamma distribution profile and predictions were made about characteristics of this nascent coffee ring such as its height and thickness under different evaporation laws. Critically, the results of the asymptotic analysis were used to determine when the assumption of a dilute solute breaks down; beyond this regime it is necessary to incorporate a model for finite concentration effects. In particular, the window over which the dilute regime is valid was found to depend strongly on whether evaporation is kinetic- or diffusion-limited.

Droplet axisymmetry greatly simplifies both the capillary flow and the solute transport problem so that significant analytical progress is possible in the asymptotic analysis. \citet{Moore2021} exploited this relative simplicity to construct consistent composite predictions for the solute mass profile. However, once axisymmetry is broken, the problem becomes much more challenging analytically and this approach is no longer feasible. 

Our aim in the present analysis is thus threefold. Firstly, we present an alternative asymptotic approach utilizing an integrated mass variable and an intermediate region that allows us to construct a composite solution for the solute concentration for an arbitrary, smooth, droplet contact set. We demonstrate the methodology in detail for a kinetic evaporation model --- in which the evaporative flux is constant --- as well as for a diffusive evaporation model for which the evaporative flux is singular at the contact line. The asymptotic results allow us to derive a similarity profile for the nascent coffee ring.  

Secondly, we use the asymptotic results to investigate the relative importance of droplet asymmetry and heterogeneity in the evaporative flux on mass flux into the contact line and the nascent coffee ring structure by considering the particular example of a droplet with an elliptical contact set. When the evaporation is dominated by diffusion, we corroborate the findings of \citet{Saenz2017} by demonstrating that there is an increased mass flux of solute into the contact line along the highly-curved semi-major axis of the ellipse compared to an axisymmetric droplet of the same volume and contact line length. However, when considering the constant evaporative flux model, we demonstrate that asymmetry in the droplet profile alone results in the same behaviour. In both cases, we derive analytical expressions for the nascent coffee ring profile, as well as key characteristics such as its height and width.

Importantly, we also demonstrate that the increased mass flux does not necessarily translate to a higher coffee ring peak. Indeed, for a diffusively-evaporating droplet, we show that, as the eccentricity of the elliptical contact set is changed but the droplet volume and perimeter are fixed, the coffee-ring profile along the semi-major axis undergoes a transition from a thinner, higher coffee ring than that for the corresponding axisymmetric droplet to a lower, shallower coffee ring. On the other hand, for a uniformly-evaporating droplet, the coffee-ring peak along the semi-major axis is always larger than that for the equivalent circular droplet, although there is still a transition from a narrower to a thicker ring.

Finally, we are able to show that this enhanced coffee-ring effect leads to a reduction in the applicability of the dilute model along the semi-major axis, although the duration of applicability is correspondingly longer along the semi-minor axis, where the coffee-ring effect is weaker.

The content of the paper is as follows. In \textsection \ref{sec:ProblemConfig}, we present a mathematical model for the evaporation of a thin droplet with an arbitrary, smooth contact set containing a dilute solute, describing the liquid flow in the droplet in the surface tension-dominated limit. We then consider the solute transport problem in detail in \textsection \ref{sec:Asymptotics}, focussing on the physically-relevant regime in which advection dominates diffusion except in a boundary layer close to the pinned contact line. We use our asymptotic analysis to investigate the evaporation of droplets with an elliptical contact set in \textsection \ref{sec:Elliptical}, uncovering the role played by droplet geometry in the formation of the nascent coffee-ring, as well as exploring the limitations of the dilute model. We conclude by summarizing our results in \textsection \ref{sec:Summary}, as well as discussing possible applications and extensions of the model.


\section{Formulation of the mathematical model} \label{sec:ProblemConfig}

A droplet of liquid of volume $V^{*}$ lies on a planar substrate at $z^{*} = 0$, where we take Cartesian axes $(x^{*},y^{*},z^{*})$ centred with origin inside the droplet contact set $\Omega^{*}$. Here and hereafter, an asterisk denotes a dimensional variable. The droplet contact line is denoted by $\partial\Omega^{*}$ and, throughout our analysis, we shall assume that it is \emph{pinned}; this is a reasonable assumption for the majority of the drying time \cite[see][]{Hu2002}. We shall assume that the droplet is \textit{thin} so that if we denote a typical size of the contact set by $R^{*}$, we have $\delta = V^{*}/R^{*3}\ll1$. We note that in our analysis of elliptical contact sets in \textsection \ref{sec:Elliptical}, we shall take $R^{*}$ to be the length of the semi-minor axis, so that when comparing droplets of the same initial volume and perimeter, $\delta$ will necessarily change.

The incompressible Newtonian liquid has density $\rho^{*}$ and viscosity $\mu^{*}$, while the air-liquid surface traction is taken to be due to a constant surface tension with coefficient denoted by $\sigma^{*}$. We shall assume that the droplet is sufficiently small that the effects of gravity are negligible, i.e. we assume that the Bond number $\mbox{Bo} = \rho^{*} g^{*} R^{*2}/\sigma^{*}$ is small, where $g^{*}$ is the gravitational acceleration. The liquid lies in the region $0<z^{*}<h^{*}(x^{*},y^{*},t^{*})$ for $(x^{*},y^{*})\in\Omega^{*}$, where the air-liquid interface lies at $z^{*} = h^{*}(x^{*},y^{*},t^{*})$. The liquid velocity and pressure are denoted by $\mbf{u}^{*}(x^{*},y^{*},z^{*},t^{*})$ and $p^{*}(x^{*},y^{*},z^{*},t^{*})$, respectively.

The liquid contains a non-volatile solute of concentration $\phi^{*}(x^{*},y^{*},z^{*},t^{*})$, which is initially uniformly-distributed, with concentration $\phi_{\mathrm{init}}^{*}$. One of our key assumptions is that the solute is sufficiently dilute that the flow within the drop is unaffected by its presence. Under the dilute assumption, we can decouple the flow in the liquid drop from the solute transport. One of our aims is to test the validity of this assumption local to the pinned contact line, where the solute concentration increases as a result of the growth of the nascent coffee ring.

The droplet evaporates into the surrounding air, which induces a flux of vapour $E^{*}$ at the droplet surface. The evaporative process is assumed to be quasi-steady, which is reasonable for a wide range of different liquid-substrate combinations \cite[][]{Hu2002}. The evaporative flux, $E^{*}$, combined with the geometry of the droplet drives a liquid flow of typical size $U^{*} = \mathcal{E}^{*}/\delta\rho^{*}$ towards the pinned contact line, where $\mathcal{E}^{*}$ is a typical size of the evaporative flux, which depends upon the evaporation model chosen \cite[see, for example,][]{Murisic2011}.


\subsection{Dimensionless model}

\begin{figure}
\centering \scalebox{0.75}{\epsfig{file=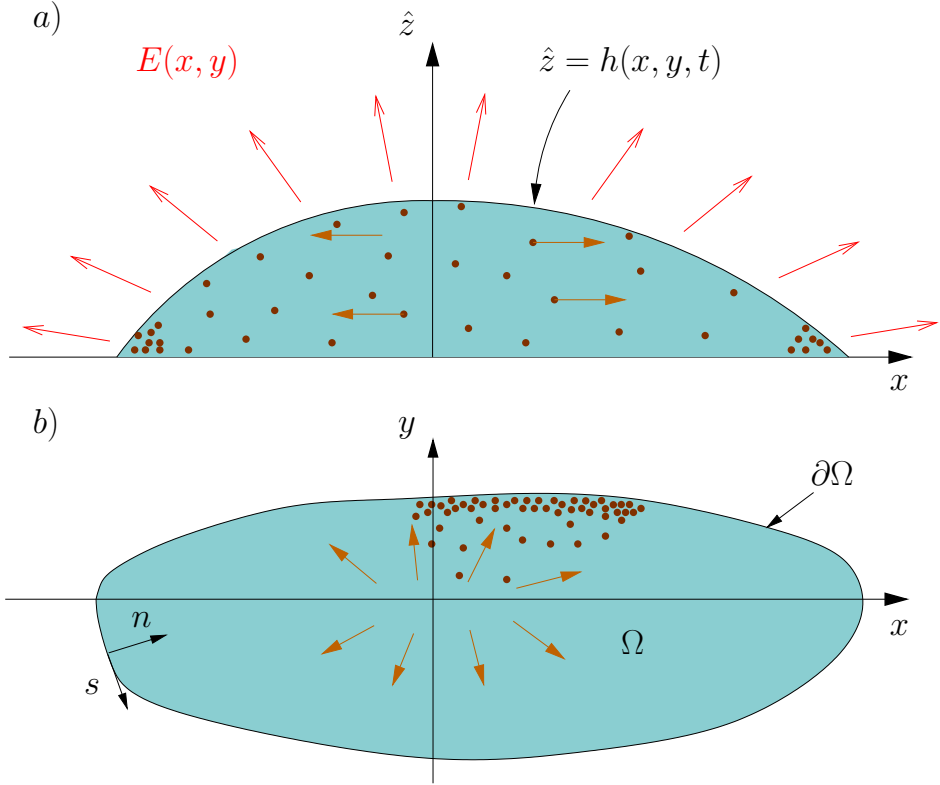}}
\caption{$a)$ Side-on and $b)$ top-down view of a partially-wetting liquid droplet with dimensionless contact set $\Omega$ and pinned contact line $\partial\Omega$ in the plane $\hat{z} = 0$ of the rigid impermeable substrate. The droplet evaporates into the surrounding air with flux $E(x,y)$ as indicated by the red arrows. The orange arrows indicate the evaporation-induced solute velocity. The liquid-air interface is denoted by $\hat{z} = h(x,y,t)$. The local orthogonal curvilinear coordinate system $(s,n)$ embedded in the contact line described in \textsection \ref{sec:Coordinates} can be seen in $b)$.}
\label{fig:DropletConfig} 
\end{figure}

The derivation of the model for the liquid flow and solute transfer are a direct extension of the axisymmetric case presented in \citet{Moore2021}, so here we shall present the model directly in dimensionless form in the interests of brevity. Exploiting the thinness of the droplet and assuming that the flow is sufficiently slow to be dominated by viscosity, the pertinent scalings are
\begin{linenomath}
 \begin{equation}
  (x^{*},y^{*}) =  R^{*}(x,y), \quad z^{*} = 
  \begin{cases}
   \delta R^{*}\hat{z} & \mbox{in the droplet},\\
   R^{*}z & \mbox{in the air},
  \end{cases} \quad t^{*} = \frac{R^{*}t_{f}}{U^{*}}t,
 \end{equation}
\end{linenomath}
and
\begin{linenomath}
 \begin{alignat}{1}
 & \quad  h^{*} = \delta Rh, \quad \mbf{u}^{*} = (u^{*},v^{*},w^{*}) = U^{*}\left(u,v,\delta w\right), \quad  \nonumber \\
  & \, p^{*} = p_{\mathrm{atm}}^{*} + \frac{\mu^{*}U^{*}}{\delta^{2}R^{*}}p, \quad E^{*} = \mathcal{E}^{*}E, \quad \phi^{*} = \phi_{\mathrm{init}}^{*}\phi, 
 \end{alignat}
\end{linenomath}
where $p_{\mathrm{atm}}^{*}$ is the atmospheric pressure and the (dimensionless) dryout time of the droplet is defined by
\begin{linenomath}
 \begin{equation}
  t_{f} = \left(\iint_{\Omega}E\,\mbox{d}S\right)^{-1}.
  \label{eqn:EvaporationTimeNonDim}
 \end{equation}
\end{linenomath}
Note that we have scaled time such that the dimensionless lifetime of the droplet is $0<t<1$ to simplify the analysis going forward. The dimensionless droplet configuration is shown in figure \ref{fig:DropletConfig}.

\subsubsection{Flow model}

As described in, for example, \citet{Deegan2000,FreedBrown2015, Moore2021}, since the droplet is assumed to be thin, to leading order the fluid flows within the drop according to the lubrication equations given by
\refstepcounter{equation}
\begin{linenomath}
 $$
  \frac{1}{t_{f}}\frac{\partial h}{\partial t} + \nabla\cdot\left(h\bar{\mbf{u}}\right) = - E, \quad \bar{\mbf{u}} = -\frac{h^{2}}{3}\nabla p, \quad p = -\frac{1}{\Ca}\nabla^{2}h
  \eqno{(\theequation{\mathit{a},\mathit{b},\mathit{c}})}
  \label{eqn:FlowNonDim1}
 $$
\end{linenomath}
for $(x,y)\in\Omega$, $0<t<1$, where $\nabla = (\partial/\partial x,\partial/\partial y)^{T}$. Here, $\bar{\mbf{u}}(x,y,t)$ is the leading-order depth-averaged velocity of the droplet, the leading-order pressure $p(x,y,t)$ is independent of $\hat{z}$ and the capillary number is 
\begin{linenomath}
 \begin{equation}
  \Ca = \frac{U^{*}\mu^{*}}{\sigma^{*}\delta^{3}}.
 \end{equation}
\end{linenomath}
Equations (\ref{eqn:FlowNonDim1}) are supplemented by boundary conditions that require the droplet thickness to vanish and there to be no flux of liquid through the pinned contact line, so that
\refstepcounter{equation}
\begin{linenomath}
$$
 h = h\bar{\mbf{u}}\cdot\mbf{n} = 0 \quad \mathrm{on} \quad \partial\Omega,
\eqno{(\theequation{\mathit{a},\mathit{b}})}
 \label{eqn:FlowNonDim2}
$$
\end{linenomath}
where $\mbf{n}$ is the outward pointing unit normal to $\partial\Omega$, while we must prescribe the initial droplet profile, $h_{0}$ say, so that
\begin{linenomath}
 \begin{equation}
  h(x,y,0) = h_{0}(x,y) \quad \mathrm{for} \quad (x,y) \in \Omega. 
  \label{eqn:FlowNonDim3}
 \end{equation}
\end{linenomath}

We note that, in the absence of suitable regularization (such as imposing a Navier slip condition, rather than no slip on the substrate), a local analysis \cite[along the lines of that described in, for example,][]{Saxton2016} reveals that the contact angle is unbounded at the contact line for both the kinetic and diffusive models. Fortunately, that these effects are localized and so we neglect them in the present analysis.

\subsubsection{Solute model}

As described in, for example, \citet{Wray2014,Pham2017,Moore2021}, for a thin droplet, the solute concentration $\phi$ is independent of $\hat{z}$ to leading order and hence satisfies the advection-diffusion equation given by
\begin{linenomath}
 \begin{equation}
   \frac{1}{t_{f}}\frac{\partial}{\partial t}\left(h\phi\right) + \nabla\cdot\left(h\phi\bar{\mbf{u}} - \frac{1}{\Pe}h\nabla\phi\right) = 0
 \label{eqn:SoluteNonDim1}
 \end{equation}
\end{linenomath}
for $(x,y)\in\Omega$ and $0<t<1$, which is to be solved subject to the boundary condition of no solute flux through the contact line, given by
\begin{linenomath}
 \begin{equation}
  \left(h\phi\bar{\mbf{u}} - \frac{1}{\Pe}h\nabla\phi\right)\cdot\mbf{n} = 0 \quad \mathrm{for} \quad (x,y) \in \partial\Omega,
  \label{eqn:SoluteNonDim2}
 \end{equation}
\end{linenomath}
together with the initial condition
\begin{linenomath}
 \begin{equation}
  \phi(x,y,0) = 1 \quad \mathrm{for} \quad (x,y)\in\Omega.
  \label{eqn:SoluteNonDim3}
 \end{equation}
\end{linenomath}
The relative importance of advection and diffusion is governed by the solutal P\'{e}clet number
\begin{linenomath}
 \begin{equation}
 \Pe = \frac{U^{*}R^{*}}{D_{\phi}^{*}},
 \end{equation}
\end{linenomath}
where $D_{\phi}^{*}$ is the solutal diffusion coefficient. We note that the reduced model is valid so long as $\delta^{2}\Pe\ll1$. For thicker droplets with $\delta^{2}\Pe$ of order unity or larger, other effects such as the capture and transport of solute along the droplet free surface may be relevant \cite[][]{Kang2016}.

\subsubsection{Evaporation models} \label{sec:EvaporativeModel}

The appropriate evaporation model depends on the chemical and thermal properties of the liquid and substrate, as well as properties of the surrounding air, such as the ambient humidity and the ease with which vapour can be transported away from the droplet. It is not the purpose of this paper to determine what is the most appropriate evaporation model, but rather to investigate the formation of the nascent coffee ring in different regimes. For this reason, we shall concentrate on two illustrative cases by considering the \textit{kinetic} and \textit{diffusive} evaporative models, as in \citet{Moore2021}.

As described in detail by \citet{Murisic2011}, in a kinetic evaporation model, evaporation is limited by the liquid phase alone and the dimensionless evaporative flux in such cases is well approximated by $E = 1/(1 + h/\mathcal{K})$, where $\mathcal{K}$ is a parameter that depends on the thermodynamic properties of the system. For water evaporating on silicon, \citet{Murisic2011} find that $\mathcal{K} \approx 10$. We can therefore reasonably take the evaporative flux to be uniform across the drop and we therefore set
\begin{linenomath}
 \begin{equation}
  E = 1 \; \mbox{for all} \; (x,y)\in\Omega.
  \label{eqn:KineticFluxNonDim}
 \end{equation}
\end{linenomath}
We note that a constant evaporative flux may also occur in other situations where the evaporation model is not diffusion-limited, for example for water droplets evaporating on a hydrogel bath \cite[][]{Boulogne2016}.

In a diffusive evaporative model, evaporation is limited by the transport of the liquid vapour away from the air-liquid interface. Following, for example, \citet{Deegan2000} and \citet{Hu2002}, we assume that this process is quasi-steady and dominated by diffusion, so that the dimensionless vapour concentration $c$ satisfies the mixed boundary value problem
\begin{linenomath}
\begin{equation}
\begin{aligned}
  \displaystyle{\nabla^{2}c} &= 0 \quad \mbox{in} \quad z>0,  \\[5pt]
  \displaystyle{c(x,y,0)} &= 1 \quad \mbox{for} \quad (x,y) \in \Omega,  \\
  \displaystyle{\frac{\partial c}{\partial z}(x,y,0)} &= 0 \quad \mbox{for} \quad (x,y)\notin \Omega,  \\[1pt]
  \displaystyle{c} &\rightarrow 0 \quad \mbox{as} \quad x^{2}+y^{2}+z^{2}\rightarrow\infty.
  \label{eqn:DiffusionVapour}
 \end{aligned}
\end{equation}
\end{linenomath}
Here we have used the fact that the droplet is thin to linearize the boundary conditions onto the substrate, $z = 0$. The dimensionless evaporative flux is then given by
\begin{linenomath}
 \begin{equation}
 E = -\frac{\partial c}{\partial z}(x,y,0) \quad \mbox{for} \quad (x,y)\in\Omega.
 \label{eqn:DiffusiveFluxNonDim}
 \end{equation}
\end{linenomath}

\subsection{Small-capillary number flow}

As discussed in, for example, \citet{Deegan2000, FreedBrown2015, Moore2021}, the pertinent physical limit of interest in a wide range of problems is that in which the capillary number is small. This is due to the fact that the timescale for evaporation is typically many orders of magnitude larger than the timescale for capillary relaxation. As $\Ca\rightarrow0$, it follows from (\ref{eqn:FlowNonDim1})--(\ref{eqn:FlowNonDim3}) that the leading-order flow is given by 
\begin{linenomath}
 \begin{alignat}{2}
  p(x,y,t) & \, \sim && \, \frac{\alpha}{\Ca}(1-t) + \frac{1}{\alpha^{3}(1-t)^{3}}P(x,y), \label{eqn:SmallCaSol1}\\[2mm]
  h(x,y,t) & \, \sim && \, \alpha(1-t)H(x,y), \label{eqn:SmallCaSol2}\\[2mm]
  \bar{\mbf{u}}(x,y,t) & \, \sim && \, -\frac{H^{2}}{3\alpha(1-t)}\nabla P, \label{eqn:SmallCaSol3}
 \end{alignat}
\end{linenomath}
where the constant
\begin{linenomath}
 \begin{equation}
  \alpha = \left(\iint_{\Omega}H\,\mbox{d}S\right)^{-1}
  \label{eqn:alpha}
 \end{equation}
\end{linenomath}
ensures that the initial dimensionless droplet volume is unity. The function $H(x,y)$ is found from (\ref{eqn:FlowNonDim1}c) and (\ref{eqn:FlowNonDim2}b) to satisfy the Dirichlet problem given by
\refstepcounter{equation}
\begin{linenomath}
$$
  \nabla^{2}H = -1 \quad \mathrm{in} \quad \Omega, \quad H = 0 \quad \mathrm{on} \quad \partial\Omega,
\eqno{(\theequation{\mathit{a},\mathit{b}})}
\label{eqn:Poisson}
$$
\end{linenomath}
while by (\ref{eqn:FlowNonDim1}a,b) and (\ref{eqn:FlowNonDim2}a), the pressure perturbation $P(x,y)$ satisfies the Neumann problem
\refstepcounter{equation}
\begin{linenomath}
 $$
 \nabla\cdot\left(-\frac{H^{3}}{3}\nabla P\right) = -E + \frac{\alpha}{t_{f}}H \quad \mbox{in} \quad \Omega, \quad -\frac{H^{3}}{3}\nabla P\cdot\mbf{n} = 0 \quad \mbox{on} \quad \partial\Omega.
 \eqno{(\theequation{\mathit{a},\mathit{b}})}
\label{eqn:PressurePert}
 $$
\end{linenomath}
We note that the solvability condition for the latter problem is automatically satisfied because of the definitions of $t_{f}$ and $\alpha$ given by (\ref{eqn:EvaporationTimeNonDim}) and (\ref{eqn:alpha}), respectively.

In the surface tension-dominated limit, the initial free surface profile, $h_{0}(x,y)$ is unused. In reality, there is a short time scale over which an arbitrary profile $h_{0}(x,y)$ relaxes under surface tension to the form given by (\ref{eqn:SmallCaSol2}) \cite[][]{Lacey1982, deGennes1985, Oliver2015}. This process happens on the timescale for capillary action, namely $t = O(\Ca)$, and therefore extremely quickly compared to the evaporation-induced flow studied here. We shall therefore neglect its effects in our analysis.

Typically, (\ref{eqn:Poisson}) and (\ref{eqn:PressurePert}) will need to be solved numerically to recover the leading-order flow in the droplet given by (\ref{eqn:SmallCaSol3}), although some analytical progress may be made in simple geometries. In particular, \citet{Deegan2000, Popov2005, Moore2021} discuss the axisymmetric problem in detail. It is also worth noting that the velocity profile will necessarily include a stagnation point in the droplet interior. In particular, in the late stages of coffee ring formation, the flow close to the stagnation point is the driving factor in the characteristic `fadeout' of the coffee-ring profile \cite[][]{Witten2009}.

\subsection{Formulation in terms of contact line coordinates} \label{sec:Coordinates}

Once the flow has been determined, the solute concentration $\phi$ is determined from (\ref{eqn:SoluteNonDim1})--(\ref{eqn:SoluteNonDim3}). As discussed in \citet{Moore2021}, the pertinent asymptotic limit for a wide range of real-world evaporation problems --- including for example, the experiments and simulations of \citet{Kajiya2008} and \citet{Saenz2017} --- is that in which the P\'{e}clet number is large, so that
\begin{linenomath}
 \begin{equation}
   \ve:=1/\Pe\ll1.
 \end{equation}
\end{linenomath}
In this regime, solute advection dominates in the bulk of the droplet, increasing the concentration local to the contact line and driving there a competing diffusive flux. 

We pursue this matched asymptotic analysis in detail in \textsection \ref{sec:Asymptotics}, but to do so it is expedient to introduce a local orthogonal curvilinear coordinate system, $(s,n)$, embedded in the contact line geometry, as illustrated in figure \ref{fig:DropletConfig}b. The normal direction, $n$, points into the droplet and the arc length, $s$, is measured anticlockwise around the contact line, which has curvature $\kappa(s)$, which here we take to be negative if the centre of the osculating circle lies to the left of the contact line as it is traversed in the anticlockwise direction (e.g. $\kappa = -1$ for a circular contact set of unit radius). We also denote the $s$- and $n$-components of the depth-averaged velocity by $\bar{u}_{s}$ and $\bar{u}_{n}$ respectively.

Such a coordinate system is well-defined provided that $\partial\Omega$ is sufficiently smooth and that there are no vanishingly small neck regions in the droplet footprint. We shall assume that not only are these conditions met but, further, that the region over which $(s,n)$ are well-defined extends much further than an $O(\ve)$-distance from the contact line. (Such conditions will be met in the vast majority of physically-relevant scenarios, so this is not a particularly restrictive assumption.) For situations in which a sharp corner exists on the boundary, \citet{Popov2003} have performed a local analysis that reveals the underlying scaling laws associated with the coffee ring growth.

\subsubsection{Local behaviour of the flow model}

We can use the contact line coordinate system to determine the local behaviour of the droplet free surface profile, evaporative flux and the liquid flow. These results will be of vital importance in determining the correct asymptotic structure as $\ve\rightarrow0$ in \textsection \ref{sec:Asymptotics}. We introduce the contact angle 
\begin{linenomath}
 \begin{equation}
  \theta_{c}(s,t) = \lim_{n\rightarrow0^{+}}\frac{\partial h}{\partial n}(s,n,t),
  \label{eqn:ContactAngle}
 \end{equation}
\end{linenomath}
which we assume to exist and be nonzero for $0<t<1$, so that the droplet contact angle decays linearly with time with $\theta_{c}(s,t) = (1-t)\psi(s)$ according to (\ref{eqn:SmallCaSol2}). The contact angle acts as a degree of freedom in the sense that it is determined globally rather than locally. We note that we have abused notation in the sense that the contact angle introduced here is for the leading-order droplet profile in the small-capillary number limit, as governed by (\ref{eqn:Poisson}), rather than that for the original model, governed by (\ref{eqn:FlowNonDim1}), (\ref{eqn:FlowNonDim2}) and (\ref{eqn:FlowNonDim3}) --- the latter does not exist in the absence of regularization, as mentioned above.

The local behaviour of the liquid flow depends strongly on the chosen evaporative model. For a kinetic evaporative flux, with $E = 1$, a local analysis of (\ref{eqn:SmallCaSol3}), (\ref{eqn:PressurePert}) and using the fact that $h\sim\theta_{c}(s,t)n$ as $n\rightarrow0^{+}$, implies that, close to the contact line
\begin{linenomath}
 \begin{equation}
   \bar{u}_{s} \sim \frac{-3n}{\theta_{c}(s,t)^{2}}\frac{\partial}{\partial s}\theta_{c}(s,t), \; \bar{u}_{n} \sim \frac{-1}{\theta_{c}(s,t)} \;\mbox{as} \; n\rightarrow0^{+}.
  \label{eqn:Un_local}
 \end{equation}
\end{linenomath}

For a diffusive evaporative flux, the change from Dirichlet to Neumann boundary condition in (\ref{eqn:DiffusionVapour}) dictates that
\begin{linenomath}
 \begin{equation}
  E(s,n) \sim \frac{\chi(s)}{\sqrt{n}} \quad \mbox{as} \quad n\rightarrow0^{+},
  \label{eqn:DiffEvapLocal}
 \end{equation}
\end{linenomath}
where $\chi(s)$ is a degree of freedom. Then, by a similar analysis, we can show that the local behaviour of the liquid velocity at the contact line is given by
\begin{linenomath}
 \begin{equation}
  \bar{u}_{s} \sim \frac{4\theta_{c}(s,t)^{2}}{3}\frac{\partial}{\partial s}\left(\frac{\chi(s)}{\theta_{c}(s,t)^{3}}\right)\sqrt{n}, \quad \bar{u}_{n} \sim \frac{-2\chi(s)}{\theta_{c}(s,t)}\frac{1}{\sqrt{n}} \quad \mbox{as} \quad n\rightarrow0^{+}. \label{eqn:Un_local_diffusive}
 \end{equation}
\end{linenomath}

It is clear that the singularity in the diffusive evaporative flux drives a stronger normal velocity close to the contact line than the kinetic evaporative flux. Moreover, while in both cases the local velocity behaviour depends on the shape of the free boundary at the contact line --- with the degree of freedom $\theta_{c}(s,t)$ being independent of the evaporation model --- the coefficient of the normal velocity singularity in the diffusive regime also depends on the coefficient $\chi(s)$ of the evaporative flux singularity. The latter behaviour was noted by \citet{Saenz2017} to have a strong effect on the local coffee-ring profile.

\subsubsection{Local formulation of the solute problem and the integrated mass variable}

In the region in which the $(s,n)$ coordinate system is well-defined, the solute transport equation (\ref{eqn:SoluteNonDim1}) can be written as
\begin{linenomath}
 \begin{eqnarray}
  \frac{1}{t_{f}}\frac{\partial}{\partial t}\left(a h\phi\right)  & + & \frac{\partial}{\partial s}\left(h\phi\bar{u}_{s} - \frac{\ve h}{a}\frac{\partial\phi}{\partial s}\right) +  \frac{\partial}{\partial n}\left(ah\phi\bar{u}_{n}-\ve ah\frac{\partial\phi}{\partial n}\right) = 0,
  \label{eqn:LocalSoluteNonDim1}
 \end{eqnarray}
\end{linenomath}
where the scale factor $a = 1+\kappa(s)n$, while the no-flux boundary condition (\ref{eqn:SoluteNonDim2}) becomes
\begin{linenomath}
 \begin{equation}
  h\phi\bar{u}_{n}-\ve h\frac{\partial\phi}{\partial n} = 0 \; \mbox{on} \; n = 0. 
  \label{eqn:LocalSoluteNonDim2}
 \end{equation}
\end{linenomath}

In order to facilitate our matched asymptotic analysis, we shall find it convenient to introduce the \emph{integrated mass variable}, defined by
\begin{linenomath}
\begin{equation}
 \mathcal{M}(s,n,t) = \int_{0}^{n}a(s,\nu)\phi(s,\nu,t)h(s,\nu,t)\,\mbox{d}\nu.
\label{eqn:WeightedMassVariable}
\end{equation}
\end{linenomath}
If we integrate (\ref{eqn:LocalSoluteNonDim1}) in the normal direction from $0$ to $n$ and apply (\ref{eqn:LocalSoluteNonDim2}), we see that the integrated mass variable satisfies
\begin{linenomath}
\begin{eqnarray}
 \frac{1}{t_{f}}\frac{\partial\mathcal{M}}{\partial t} + \frac{\partial}{\partial s}\int_{0}^{n} & &h\phi\bar{u}_{s} - \frac{\ve h}{a}\frac{\partial\phi}{\partial s}\,\mbox{d}\nu + ah\phi\bar{u}_{n} - \ve ah\frac{\partial\phi}{\partial n}  = 0, \label{eqn:LocalIMV1}
\end{eqnarray}
\end{linenomath}
while by the initial condition (\ref{eqn:SoluteNonDim3}), we have
\begin{linenomath}
 \begin{equation}
  \mathcal{M}(s,n,0) = \alpha \int_{0}^{n}a(s,\nu)H(s,\nu)\,\mbox{d}\nu, \label{eqn:LocalIMV2}
 \end{equation}
\end{linenomath}
where $\alpha$ and $H$ are given by (\ref{eqn:alpha}) and (\ref{eqn:Poisson}), respectively.

\subsection{Summary}

In summary, given a model for the evaporative flux $E$, such as (\ref{eqn:KineticFluxNonDim}) or (\ref{eqn:DiffusiveFluxNonDim}), we first determine at leading order in $\Ca\ll1$ the dryout time, $t_{f}$, the free surface profile, $h$, the liquid pressure, $p$, and liquid velocity $\bar{\mbf{u}}$ from (\ref{eqn:EvaporationTimeNonDim}), (\ref{eqn:SmallCaSol1})--(\ref{eqn:SmallCaSol3}), (\ref{eqn:alpha}), (\ref{eqn:Poisson}) and (\ref{eqn:PressurePert}). The local behaviour of the leading-order free surface profile at the contact line is given by (\ref{eqn:ContactAngle}), while the liquid velocity has local behaviour given by (\ref{eqn:Un_local}) in the kinetic evaporative regime and by (\ref{eqn:Un_local_diffusive}) in the diffusive evaporative regime.

Once the leading-order free surface and the fluid velocity have been found, the solute concentration, $\phi$, may be determined by solving (\ref{eqn:SoluteNonDim1})--(\ref{eqn:SoluteNonDim3}), which in the region of validity of the contact line-based coordinate system imply (\ref{eqn:LocalSoluteNonDim1})--(\ref{eqn:LocalSoluteNonDim2}) for $\phi$ and hence (\ref{eqn:LocalIMV1})--(\ref{eqn:LocalIMV2}) for the integrated mass variable defined in (\ref{eqn:WeightedMassVariable}). We shall exploit both forms of the solute concentration problem in our upcoming asymptotic analysis.


\section{Asymptotic solution of the advection-dominated limit} \label{sec:Asymptotics}

We seek to determine how the interplay between advection and diffusion in (\ref{eqn:SoluteNonDim1})--(\ref{eqn:SoluteNonDim3}) drives the growth of the nascent coffee ring for $\Pe\gg1$ ($\ve\ll1$), in which case solute transport is dominated by advection in the outer region away from the contact line \cite[][]{Deegan1997,Deegan2000,Popov2005,Witten2009,Moore2021}. Close to the contact line, where the solute concentration rapidly builds up and large concentration gradients form, the advective transport is balanced by diffusion in an inner region. As described by \citet{Moore2021}, the asymptotic analysis is strongly-dependent on the choice of evaporative model. We give details of the analysis in \textsection\textsection \ref{sec:Outer}--\ref{sec:Similarity} for the kinetic evaporative flux (\ref{eqn:KineticFluxNonDim}) and summarize the corresponding results for the diffusive evaporative flux (\ref{eqn:DiffusiveFluxNonDim}) in \textsection \ref{sec:Diffusive}.

\subsection{Outer region} \label{sec:Outer}

In the bulk of the drop, we expand $\phi \sim \phi_{0}$ as $\ve\rightarrow0$. At leading order, we recover from (\ref{eqn:SoluteNonDim1}) and (\ref{eqn:SoluteNonDim3}) the initial value problem
\begin{linenomath}
 \begin{equation}
  \frac{1}{t_{f}}\frac{\partial}{\partial t}\left(h\phi_{0}\right) + \nabla\cdot\left(h\bar{\mbf{u}}\phi_{0}\right) = 0 \quad \mathrm{in} \quad \Omega, \quad \phi_{0}(x,y,0) = 1 \quad \mbox{in} \quad \Omega.
  \label{eqn:SoluteOuter}
 \end{equation}
\end{linenomath}
Thus, the leading-order behaviour in the bulk is simply to advect solute towards the contact line \cite[see, for example,][]{Deegan2000,Popov2005,Witten2009}, as expected in the large-solutal P\'{e}clet number limit. The solution of (\ref{eqn:SoluteOuter}) may be written in the form
\begin{linenomath}
 \begin{equation}
  (h\phi_{0})(x,y,t) = \frac{\alpha H(X,Y)}{J(X,Y,t)}
  \label{eqn:OuterSolution}
 \end{equation}
\end{linenomath}
where $(X,Y)$ is the initial location of the fluid element positioned at $(x,y)$ at time $t$ and $J(X,Y,t)$ is the Jacobian of the Eulerian--Lagrangian transformation.

Hence, in the dilute model, the solute in the bulk simply follows streamlines to the contact line, as discussed previously by, for example, \citet{Deegan2000} and \citet{Witten2009} --- even though the flow is unsteady here, the particle paths coincide with the streamlines by virtue of the separable nature of the time dependence in (\ref{eqn:SmallCaSol3}). What is particularly useful about the form of the solution (\ref{eqn:OuterSolution}) is that, once we have calculated $H$ and $\bar{\mbf{u}}$, rather than solving the hyperbolic problem (\ref{eqn:SoluteOuter}), we can instead solve for the Jacobian by integrating Euler's identity,
\begin{linenomath}
\begin{equation}
 \frac{\mbox{D}}{\mbox{D}t}(\log{J}) = t_{f}\nabla\cdot\bar{\mbf{u}}, \label{eqn:EulersIdentity}
\end{equation}
\end{linenomath}
along a streamline from $(X,Y)$, treating 
(\ref{eqn:EulersIdentity}) as a first-order ODE with the initial condition 
$J(X,Y,0) = 1$. Further, we can avoid calculating the divergence numerically by utilizing (\ref{eqn:SmallCaSol3}) and (\ref{eqn:PressurePert}). This methodology lends itself particularly well to geometries in which the 
Poisson problem (\ref{eqn:Poisson}) is solvable analytically, as in the example of an elliptical contact set that we consider in \textsection \ref{sec:Elliptical}. We discuss the numerical treatment of (\ref{eqn:EulersIdentity}) further in Appendix \ref{appendix:Numerics}.

We note that, according to (\ref{eqn:LocalIMV1}), the integrated mass variable $\mathcal{M} \sim \mathcal{M}_{0}$ as $\ve\rightarrow0$, where $\mathcal{M}_{0}$ evolves according to
\begin{linenomath}
 \begin{equation}
 \frac{1}{t_{f}}\frac{\partial\mathcal{M}_{0}}{\partial t} + \frac{\partial}{\partial s}\int_{0}^{n}h\phi_{0}\bar{u}_{s}(s,\nu,t)\,\mbox{d}\nu + (1+\kappa n)h\phi_{0}\bar{u}_{n} = 0.
 \end{equation}
\end{linenomath}
Taking the limit $n\rightarrow0$, we deduce that
\begin{linenomath}
 \begin{equation}
  \mathcal{M}_{0}(s,0^{+},t) = -t_{f}\int_{0}^{t}(h\phi_{0}\bar{u}_{n})(s,0^{+},\tau)\,\mbox{d}\tau,\label{eqn:MassFluxIntoBL}
\end{equation}
\end{linenomath}
which is simply the leading-order accumulated mass flux that has flowed into the contact line from the outer region up to time $t$. This quantity will be essential later in the matching between the outer and inner regions.

That an inner region is necessary is evident by considering the local behaviour of the solute concentration at the contact line. Upon recalling (\ref{eqn:ContactAngle}) and (\ref{eqn:Un_local}), we expand (\ref{eqn:SoluteOuter}) as $n\rightarrow0$ to deduce that
\begin{linenomath}
 \begin{equation}
  \phi_{0} \sim \frac{B(s,t)}{n} \quad \mbox{as} \quad n\rightarrow0,
  \label{eqn:Phi_local}
 \end{equation}
\end{linenomath}
where $B(s,t)$ is a degree of freedom that can only be determined by solving (\ref{eqn:OuterSolution})--(\ref{eqn:EulersIdentity}) for $\phi_{0}$. Hence the solute concentration is singular at the contact line, and we therefore expect large solute concentration gradients to form. Such large gradients will in turn induce a local diffusive flux, as we shall now describe. 

\subsection{Inner region} \label{sec:Inner}

To retain a leading-order balance between advection and diffusion, we introduce the inner scalings
\begin{linenomath}
 \begin{equation}
  n = \ve \hat{n}, \; h = \ve \hat{h}, \; \bar{u}_{s} = \ve \hat{u}_{s}, \; \bar{u}_{n} = \hat{u}_{n}, \; \phi = \ve^{-2}\hat{\phi}, \; \mathcal{M} = \hat{\mathcal{M}},
 \end{equation}
\end{linenomath}
where the scalings for the droplet thickness and the velocity components are determined from the local behaviours (\ref{eqn:ContactAngle}) and (\ref{eqn:Un_local}), while the scaling for $\phi$ is determined from global conservation of solute mass. 

Upon substituting these scalings into (\ref{eqn:LocalSoluteNonDim1})--(\ref{eqn:LocalSoluteNonDim2}) and expanding $\hat{\phi} \sim  \hat{\phi}_{0}$ as $\ve\rightarrow0$, to leading order we have
\begin{linenomath}
 \begin{equation}
  \frac{\partial}{\partial \hat{n}}\left(\hat{h}\hat{u}_{n}\hat{\phi}_{0} - \hat{h}\frac{\partial\hat{\phi}_{0}}{\partial\hat{n}}\right) = \frac{\partial}{\partial\hat{n}}\left(-\hat{n}\hat{\phi}_{0} - \theta_{c}(s,t)\hat{n}\frac{\partial\hat{\phi}_{0}}{\partial\hat{n}}\right) = 0
 \end{equation}
\end{linenomath}
for $n>0$, which must be solved subject to
\begin{linenomath}
 \begin{equation}
  -\hat{n}\hat{\phi}_{0} - \theta_{c}(s,t)\hat{n}\frac{\partial\hat{\phi}_{0}}{\partial\hat{n}} = 0 \; \mbox{on} \; n=0.
 \end{equation}
\end{linenomath}
Hence,
\begin{linenomath}
 \begin{equation}
  \hat{\phi}_{0} = C(s,t)\mbox{e}^{-\hat{n}/\theta_{c}(s,t)},
  \label{eqn:SoluteInner}
 \end{equation}
\end{linenomath}
where $C(s,t)$ must be determined by matching.

At this point it is clear that we cannot match na\"{i}vely between the 
leading-order-inner solution for the concentration (\ref{eqn:SoluteInner}) and the leading-order-outer given by (\ref{eqn:OuterSolution}) and (\ref{eqn:Phi_local}), since (\ref{eqn:SoluteInner}) shows that $\hat{\phi}_{0}$ decays exponentially in the far-field, while (\ref{eqn:Phi_local}) shows that $\phi_{0}=O(1/n)$ as we approach the contact line. 

It is here that we turn to the integrated mass variable, $\hat{\mathcal{M}}$, as defined by  (\ref{eqn:WeightedMassVariable}). In the inner region, the curvature term at the contact line may be neglected in the integrand in (\ref{eqn:WeightedMassVariable}), so that 
\begin{linenomath}
 \begin{equation}
  \hat{\mathcal{M}} \sim \hat{\mathcal{M}}_{0}(s,\hat{n},t) = -t_{f}\int_{0}^{\hat{n}}\theta_{c}(s,t)\nu\hat{\phi}_{0}(s,\nu,t)\,\mbox{d}\nu   
  \end{equation}
\end{linenomath}
as $\ve\rightarrow0$, with (\ref{eqn:SoluteInner}) giving
\begin{linenomath}
\begin{equation}
 \hat{\mathcal{M}}_{0} = C(s,t)\theta_{c}(s,t)^{3}\left(1 - \left(\frac{\hat{n}}{\theta_{c}(s,t)} + 1\right)\mbox{e}^{-\hat{n}/\theta_{c}(s,t)}\right).
  \label{eqn:SoluteInner_2}
 \end{equation}
\end{linenomath}
We deduce immediately from (\ref{eqn:SoluteInner_2}) that
\begin{linenomath}
 \begin{equation}
 \hat{\mathcal{M}}_{0}(s,\hat{n},t)\rightarrow C(s,t)\theta_{c}(s,t)^{3} \quad \mbox{as} \quad \hat{n}\rightarrow\infty.
 \label{eqn:WMVInnerFarField}
 \end{equation}
\end{linenomath}
Therefore, matching the integrated mass variable using (\ref{eqn:MassFluxIntoBL}) and (\ref{eqn:WMVInnerFarField}), we deduce that
\begin{linenomath}
 \begin{equation}
  C(s,t) = \frac{\mathcal{M}_{0}(s,0^{+},t)}{\theta_{c}(s,t)^{3}}.
  \label{eqn:C}
 \end{equation}
\end{linenomath}
We observe that $C(s,t)$ depends on both the accumulated mass flux $\mathcal{M}(s,0^{+},t)$ transported from the outer region into the inner region and the contact angle, $\theta_{c}(s,t)$, so that the geometry of $\Omega$ will also be a factor in determining the local solute profile.

Although we now have our leading-order-outer and leading-order-inner solutions for $\phi$, we are as yet unable to form a composite expansion for the solute mass $m = h\phi$ --- which has the advantage over the concentration $\phi$ in potential comparisons to experimental data since it is related to the absorbance of the deposit through the Beer-Lambert law \citet{Swinehart1962} --- since in the outer region $m$ is bounded but finite as we approach the contact line (cf. (\ref{eqn:ContactAngle}) and (\ref{eqn:Phi_local})), while in the inner region, $m$ decays exponentially in the far-field. \citet{Moore2021} addressed this issue for the axisymmetric case by proceeding to higher-order in the inner region. However, given the arbitrary geometry considered here, this approach is significantly more challenging. Instead, we shall construct a composite by introducing an intermediate region.

\subsection{Intermediate region} \label{sec:Intermediate}

Let us make the change of variables
\begin{linenomath}
 \begin{equation}
  n = \delta\skn{n}, \; h = \delta\skn{h}, \; \bar{u}_{s} = \delta\skn{u}_{s}, \; \bar{u}_{n} = \skn{u}_{n}, \; \phi = \ve^{-1}\skn{\phi},
  \label{eqn:Intermediate_Scalings_1}
 \end{equation}
\end{linenomath}
where $\delta(\ve)\rightarrow0$ as $\ve\rightarrow0$ and $\ve\ll\delta\ll\ve^{1/2}$. The choice of upper bound on the range of $\delta$ allows us to neglect the time derivative in (\ref{eqn:LocalSoluteNonDim1}) in the intermediate region in the analysis below. However, provided that this condition is met, we shall see that the choice of $\delta$ is arbitrary. 

Substituting the scalings (\ref{eqn:Intermediate_Scalings_1}) into (\ref{eqn:LocalSoluteNonDim1}), we deduce that 
\begin{linenomath}
 \begin{equation}
  -\frac{\partial}{\partial \skn{n}}\left(\skn{n}\skn{\phi} + \frac{\ve}{\delta}\theta_{c}(s,t)\skn{n}\frac{\partial\skn{\phi}}{\partial\skn{n}}\right) = O(\delta),
 \end{equation}
\end{linenomath}
as $\ve,\delta\rightarrow0$ with $\skn{n} = O(1)$, so that
\begin{linenomath}
 \begin{equation}
  \skn{\phi}\sim\skn{\phi}_{0} = \exp\left(-\frac{\delta\skn{n}}{\ve\theta_{c}(s,t)}\right)\left[D(s,t) + E(s,t)\mbox{Ei}\left(\frac{\delta\skn{n}}{\ve\theta_{c}(s,t)}\right)\right]
  \label{eqn:SoluteIntermediate_2}
 \end{equation}
\end{linenomath}
in the intermediate region, where $D(s,t)$ and $E(s,t)$ are unknown functions to be determined by matching and $\mbox{Ei}(x)$ is the exponential integral.

We match with the leading-order outer solution by introducing a further intermediate variable $N$, related to $n$ and $\skn{n}$ by the scalings
\begin{linenomath}
 \begin{equation}
  n = \delta^{\beta}N = \delta\skn{n} \quad (0<\beta<1).
 \end{equation}
\end{linenomath}
From the local expansion (\ref{eqn:Phi_local}) of the outer solution, we have
\begin{linenomath}
\begin{equation}
 \phi \sim \frac{B(s,t)}{\delta^{\beta} N} \; \mbox{as} \; \ve,\delta\rightarrow0\; \mbox{with} \; N=O(1). \label{eqn:Matching_1}
\end{equation}
\end{linenomath}
From the far-field expansion of the intermediate solution (\ref{eqn:SoluteIntermediate_2}), we have
\begin{linenomath}
 \begin{equation}
  \phi \sim \frac{E(s,t)\theta_{c}(s,t)}{\delta^{\beta}N} \; \mbox{as} \; \ve,\delta\rightarrow0\; \mbox{with} \; N = O(1), \label{eqn:Matching_2}
 \end{equation}
\end{linenomath}
where we have used the fact that $\mbox{Ei}(x)\sim\mbox{e}^{x}/x$ as $x\rightarrow\infty$. Matching in (\ref{eqn:Matching_1}) and (\ref{eqn:Matching_2}) gives
\begin{linenomath}
 \begin{equation}
  E(s,t) = \frac{B(s,t)}{\theta_{c}(s,t)}.
 \end{equation}
\end{linenomath}

To determine $D(s,t)$, we match with the inner solution in a similar manner. We first introduce the new intermediate variable $N$ via the scalings
\begin{linenomath}
 \begin{equation}
  n = \ve\left(\frac{\delta}{\ve}\right)^{\beta}N = \ve\hat{n} \quad (0<\beta<1).
 \end{equation}
\end{linenomath}
Then, from the leading-order inner solution given by (\ref{eqn:SoluteInner}) and (\ref{eqn:C}), we have
\begin{linenomath}
 \begin{equation}
  \phi \sim \frac{1}{\ve^{2}}\frac{\mathcal{M}_{0}(s,t)}{\theta_{c}(s,t)^{3}}\mbox{exp}\left(-\left(\frac{\delta}{\ve}\right)^{\beta}\frac{N}{\theta_{c}(s,t)}\right) \; \mbox{as} \; \ve,\delta\rightarrow0\; \mbox{with} \; N = O(1).
  \label{eqn:Matching_3}
 \end{equation}
\end{linenomath}
Meanwhile, from the near-field expansion of the intermediate solution (\ref{eqn:SoluteIntermediate_2}), we see
\begin{linenomath}
 \begin{equation}
  \phi \sim \mbox{exp}\left(-\left(\frac{\delta}{\ve}\right)^{\beta}\frac{N}{\theta_{c}(s,t)}\right)\left[\frac{D(s,t)}{\ve} + \frac{E(s,t)}{\ve}\left(\log\left(\left(\frac{\delta}{\ve}\right)^{\beta}\frac{N}{\theta_{c}(s,t)}\right) + \gamma\right)\right],
  \label{eqn:InterCLExp}
 \end{equation}
\end{linenomath}
as $\ve,\delta\rightarrow0$ with $N = O(1)$, where we have used the fact that $\mbox{Ei}(x)\sim\log{x} + \gamma$ as $x\rightarrow0$, where $\gamma$ is the Euler-Mascheroni constant. Thus, matching using (\ref{eqn:Matching_3}) and (\ref{eqn:InterCLExp}) we determine
\begin{linenomath}
 \begin{equation}
  D(s,t) = \frac{\mathcal{M}_{0}(s,t)}{\ve\theta_{c}(s,t)^{3}}.
 \end{equation}
\end{linenomath}

In summary, eliminating $\skn{n}$ in favour of $n = \delta\skn{n}$, the leading-order intermediate solution is given by
\begin{linenomath}
 \begin{equation}
  \skn{\phi} \sim \skn{\phi}_{0} = \frac{\mathcal{M}_{0}(s,t)}{\ve\theta_{c}(s,t)^{3}}\mbox{e}^{-n/\ve\theta_{c}(s,t)} + \frac{B(s,t)}{\theta_{c}(s,t)}\mbox{e}^{-n/\ve\theta_{c}(s,t)}\mbox{Ei}\left(\frac{n}{\ve\theta_{c}(s,t)}\right)
  \label{eqn:SoluteIntermediate}
  \end{equation}
\end{linenomath}
as $\ve\rightarrow0$ with $\ve \ll n \ll \ve^{1/2}$. As mentioned above, clearly the intermediate solution is independent of the particular choice of $\delta$. In particular, it allows a transition between the $1/n$ singularity in the local expansion of the leading-order-outer solute concentration (\ref{eqn:SoluteOuter}) and the exponential decay of the leading-order-inner solute concentration (\ref{eqn:SoluteInner}).

\subsection{Composite solution} \label{sec:Composite}

We can now construct an additive composite solution for the solute concentration. The composite profile is given by
\begin{linenomath}
 \begin{alignat}{2}
  \phi_{\mathrm{comp}}(x,y,t) & \, = && \, \phi_{0}(x,y,t) + \frac{1}{\ve}\frac{B(s,t)}{\theta_{c}(s,t)}\mbox{e}^{-n/\ve\theta_{c}(s,t)}\mbox{Ei}\left(\frac{n}{\ve\theta_{c}(s,t)}\right) + \frac{1}{\ve^{2}}\frac{\mathcal{M}_{0}(s,t)}{\theta_{c}(s,t)^{3}}\mbox{e}^{-n/\ve\theta_{c}(s,t)} \nonumber \\
  & \, && \, - \frac{B(s,t)}{n} - \frac{1}{\ve}\frac{B(s,t)}{\theta_{c}(s,t)}\mbox{e}^{-n/\ve\theta_{c}(s,t)}\log\left(\frac{n}{\ve\theta_{c}(s,t)}\right),
  \label{eqn:SoluteCompositeGeneral}
 \end{alignat}
\end{linenomath}
where $\phi_{0}$ is given by (\ref{eqn:SoluteOuter}). The final two terms in the first line are the intermediate and inner solutions, where we have accounted for the overlap contribution using Van Dyke's matching rule \cite[][]{VanDyke1964}. On the second line, the first term is the overlap contribution between the outer and intermediate regions, while the final term is the overlap term between the intermediate and inner solutions, which is included to ensure that the expansion is uniformly valid throughout the droplet --- if it were not included, the expansion would become unbounded at the contact line due to the logarithmic singularity in the exponential integral for small arguments. We note that the composite expansion (\ref{eqn:SoluteCompositeGeneral}) is only available where the $(s,n)$ coordinates are well-defined. While this includes the all-important neighbourhood of the contact line, it does not include the whole contact set in general. In \textsection \ref{sec:AxisymmetricValidation} and \textsection \ref{sec:MassProfiles}, we shall use the composite solution to plot transient profiles of the solute mass throughout an evaporating droplet.

\subsection{Similarity form and properties of the nascent coffee ring} \label{sec:Similarity}

We note that, in the limit $\ve\rightarrow0$, the properties of the nascent coffee ring are dominated by contributions from the leading-order-inner solution described in \textsection \ref{sec:Inner}. In particular, the leading-order-inner solute mass is given by
\begin{linenomath}
 \begin{equation}
  \hat{m}_{0}(s,\hat{n},t) = \hat{\phi}_{0}(s,\hat{n},t)\hat{h}_{0}(s,\hat{n},t) = \frac{1}{\ve}\frac{\mathcal{M}_{0}(s,0^{+},t)}{\theta_{c}(s,t)^{2}}\hat{n}\mbox{e}^{-\hat{n}/\theta_{c}(s,t)}.
  \label{eqn:InnerMass}
 \end{equation}
\end{linenomath}

Following \citet{Moore2021}, we can find a similarity form of the coffee ring profile by introducing the time-dependent modified P\'{e}clet number $\Pe_{t}$, given by
\begin{linenomath}
 \begin{equation}
  \Pe_{t} := \frac{\Pe}{1-t},
  \label{eqn:ModifiedPe}
 \end{equation}
\end{linenomath}
which measures the relative importance of advection and diffusion accounting for the time dependence of the evaporation-induced liquid velocity, which scales with $(1-t)^{-1}$, as seen in (\ref{eqn:SmallCaSol3}). Combining (\ref{eqn:InnerMass}) with (\ref{eqn:ModifiedPe}), we see that the local solute mass profile can be expressed as
\begin{linenomath}
 \begin{equation}
  \frac{\hat{m}_{0}(s,N,t)}{\Pe_{t}\mathcal{M}_{0}(s,0^{+},t)} = \frac{N}{\psi(s)^{2}}\mbox{e}^{-N/\psi(s)} = f\left(N;2,\frac{1}{\psi(s)}\right), \; N = \Pe_{t}n
  \label{eqn:SimilarityForm}
 \end{equation}
\end{linenomath}
where $f(x;k,l) = l^{k}x^{k-1}e^{-lx}/\Gamma(k)$ is the probability density function of a gamma distribution. Note that this is similar to the analysis presented for the axisymmetric droplet in \citet{Moore2021}, but with the additional dependence on the droplet geometry through $\psi(s)$ and $\mathcal{M}_{0}(s,0^{+},t)$. 

We can use the similarity form (\ref{eqn:SimilarityForm}) to estimate characteristics of the nascent coffee ring to leading order in $\ve$. In particular, the peak of the solute mass $m_{\mathrm{max}}(s,t)$ (the intensity of the coffee ring) and its location $n_{\mathrm{max}}(s,t)$ are given by
\refstepcounter{equation}
\begin{linenomath}
$$
  m_{\mathrm{max}}(s,t) = \frac{\Pe_{t}\mathcal{M}_{0}(s,0^{+},t)}{\psi(s)\mbox{e}}, \; n_{\mathrm{max}}(s,t) = \frac{\psi(s)}{\Pe_{t}}.
  \eqno{(\theequation{\mathit{a},\mathit{b}})}
  \label{eqn:MaximumAndLocation}
$$
\end{linenomath}
A measure of the radial thickness of the nascent coffee ring is given by the \textit{full-width at half-maximum}, $w_{1/2}(s)$, which is readily determined from (\ref{eqn:SimilarityForm}) to be
\begin{linenomath}
 \begin{equation}
   w_{1/2}(s,t) = n_{\mathrm{max}}(s,t)\left[W_{0}\left(-\frac{1}{2\mbox{e}}\right) - W_{-1}\left(-\frac{1}{2\mbox{e}}\right)\right],
  \label{eqn:HalfWidth}
 \end{equation}
\end{linenomath}
where $W_{0}(x)$ and $W_{-1}(x)$ are the Lambert-W functions \cite[\tit{i.e.} solutions to $w\mbox{e}^{w} = x$,][]{Olver2010}. Notably, the ring width is simply a constant fraction of the peak location. Correspondingly, when the peak is located \emph{further} from the contact line, the ring must be \emph{thicker}.

In each of (\ref{eqn:SimilarityForm})--(\ref{eqn:HalfWidth}), it is only through $\psi(s)$ and $\mathcal{M}_{0}(s,0^{+},t)$ that any dependence on $s$ may arise. To understand asymmetries in the nascent coffee ring, these functions must be understood. One of the aims of the following sections is to investigate them in detail for some specific problems.

\subsection{Asymptotic results for a diffusive evaporative flux}
\label{sec:Diffusive}

Before we move on to validate the asymptotic predictions, we now state the equivalent asymptotic results for the diffusive evaporation model in which the corresponding dimensionless evaporative flux is square-root singular at the contact line, with local expansion given by (\ref{eqn:DiffEvapLocal}). As previously, we may  neglect the effects of solute diffusion in the bulk of the droplet, so that the outer solution as described in \textsection \ref{sec:Outer} remains the same as in the kinetic evaporative model.
In particular, the local behaviour of the normal velocity (\ref{eqn:Un_local_diffusive}) means that, at the contact line,
\begin{linenomath}
 \begin{equation}
  \phi_{0} \sim \frac{B_{d}(s,t)}{\sqrt{n}} \; \mbox{as} \; n\rightarrow0,
  \label{eqn:DiffusiveOuterLocal}
 \end{equation}
\end{linenomath}
where $B_{d}(s,t)$ is a degree of freedom; we note that this is a weaker singularity than in (\ref{eqn:Phi_local}).

In addition to weakening the outer solute singularity, (\ref{eqn:Un_local_diffusive}) also necessitates a different scaling for the inner region in which the advective and diffusive fluxes balance, namely
\begin{linenomath}
 \begin{equation}
  n = \ve^{2}\hat{n}_{d}, \; h = \ve^{2}\hat{h}, \; \bar{u}_{s} = \ve\hat{u}_{s}, \; \bar{u}_{n} = \frac{1}{\ve}\hat{u}_{n}, \; \phi = \frac{1}{\ve^{4}}\hat{\phi}.
  \label{eqn:DiffusiveInnerScalings}
 \end{equation}
\end{linenomath}
Thus, the size of the inner region is an order of magnitude smaller for the diffusive evaporative flux --- $O(\ve^{2})$ compared to $O(\ve)$ --- while the solute concentration is two orders of magnitude larger --- $O(1/\ve^{4})$ compared to $O(1/\ve^{2})$. This fits with the experimentally-observed tendency for a diffusive evaporative flux to produce narrower, higher coffee rings than for a constant evaporative flux \cite[][]{Kajiya2008}. 

After substituting (\ref{eqn:DiffusiveInnerScalings}) into (\ref{eqn:SoluteNonDim1})--(\ref{eqn:SoluteNonDim3}) and expanding as $\ve\rightarrow0$, the leading-order-inner solute concentration is given by
\begin{linenomath}
 \begin{equation}
  \hat{\phi}_{0} = C_{d}(s,t)\mbox{e}^{-4\chi(s)\sqrt{\hat{n}_{d}}/\theta_{c}(s,t)},
 \end{equation}
\end{linenomath}
where the coefficient $C_{d}(s,t)$ can be determined using the integrated mass variable, $\mathcal{M}$, in a similar manner to that in which it was determined in the kinetic regime in \textsection \ref{sec:Inner}. We find that
\begin{linenomath}
 \begin{equation}
  C_{d}(s,t) = \frac{64\chi(s)^{4}\mathcal{M}_{0}(s,0^{+},t)}{3\theta_{c}(s,t)^{5}}.
 \end{equation}
\end{linenomath}

To form a composite solution for the solute concentration, we again introduce an intermediate region through the scaling
\begin{linenomath}
 \begin{equation}
  n = \delta\skn{n}_{d}, \; h = \delta\skn{h}, \; \bar{u}_{s} = \delta^{1/2}\skn{u}_{s}, \; \bar{u}_{n} = \frac{1}{\delta^{1/2}}\skn{u}_{n}, \; \phi = \frac{1}{\ve}\skn{\phi},
  \label{eqn:DiffusiveIntermediateScalings}
 \end{equation}
\end{linenomath}
where now $\ve^{2}\ll\delta\ll\ve$. Again, this range of $\delta$ is chosen so that we may neglect the time-derivative term in (\ref{eqn:LocalSoluteNonDim1}) in the intermediate region (though, again, the solution is independent of the choice of $\delta$). Pursuing a similar analysis to \textsection \ref{sec:Intermediate}, we find that the corresponding leading-order-intermediate solution is given by
\begin{linenomath}
 \begin{equation}
  \skn{\phi}\sim\skn{\phi}_{0} = \mbox{exp}\left(-\frac{\delta^{1/2}}{\ve}\frac{4\chi(s)\sqrt{\skn{n}_{d}}}{\theta_{c}(s,t)}\right)\left[D_{d}(s,t) + E_{d}(s,t)\mbox{Ei}\left(\frac{\delta^{1/2}}{\ve}\frac{4\chi(s)\sqrt{\skn{n}}}{\theta_{c}(s,t)}\right)\right],
 \end{equation}
\end{linenomath}
as $\ve,\delta\rightarrow0$ with $\skn{n}=O(1)$, where again $D_{d}(s,t)$ and $E_{d}(s,t)$ must be determined by matching. The procedure follows in a similar manner to the kinetic case in \textsection \ref{sec:Intermediate} and we find that
\begin{linenomath}
 \begin{equation}
  D_{d}(s,t) = \frac{64\chi(s)^{4}\mathcal{M}_{0}(s,0^{+},t)}{3\ve^{3}\theta_{c}(s,t)^{5}}, \; E_{d}(s,t) = \frac{4B(s,t)\chi(s)}{\theta_{c}(s,t)}.
 \end{equation}
\end{linenomath}

It follows that an additive composite expansion for the solute concentration profile is given by
\begin{linenomath}
 \begin{alignat}{2}
  \phi_{\mathrm{comp}}(x,y,t) & \, = && \, \phi_{0}(x,y,t) + \frac{1}{\ve}\frac{4\chi(s)B_{d}(s,t)}{\theta_{c}(s,t)}\mbox{e}^{-4\chi(s)\sqrt{n}/\ve\theta_{c}(s,t)}\mbox{Ei}\left(\frac{4\chi(s)\sqrt{n}}{\ve\theta_{c}(s,t)}\right) \nonumber \\
  & \, && \, + \frac{1}{\ve^{4}}\frac{64\chi(s)^{4}\mathcal{M}_{0}(s,0^{+},t)}{3\theta_{c}(s,t)^{5}}\mbox{e}^{-4\chi(s)\sqrt{n}/\ve\theta_{c}(s,t)} - \frac{B_{d}(s,t)}{\sqrt{n}} 
  \nonumber \\
  & \, && \, -\frac{1}{\ve}\frac{4\chi(s)B_{d}(s,t)}{\theta_{c}(s,t)}\mbox{e}^{-4\chi(s)\sqrt{n}/\ve\theta_{c}(s,t)}\log\left(\frac{4\chi(s)\sqrt{n}}{\ve\theta_{c}(s,t)}\right).
  \label{eqn:SoluteCompositeDiffusive}
 \end{alignat}
\end{linenomath}
As in the kinetic regime, we note that the final term is included so that the composite solution remains asymptotic throughout the whole drop.

The equivalent similarity profile for the nascent coffee ring in the diffusive evaporative regime is given by 
\begin{linenomath}
 \begin{equation}
  \frac{\hat{m}_{0}(s,N_{d},t)}{\Pe_{t}^{2}\mathcal{M}_{0}(s,0^{+},t)} = \frac{2\chi(s)}{3\psi(s)}f\left(\sqrt{N_{d}},3,\frac{4\chi(s)}{\psi(s)}\right), \quad N_{d} = \Pe_{t}^{2}n
  \label{eqn:Similarity_Diff}
 \end{equation}
\end{linenomath}
where the shape function of the gamma distribution is now $3$ compared to $2$ in the kinetic regime. The coffee ring peak $m_{\mathrm{max}}(s,t)$ and its location $n_{\mathrm{max}}(s,t)$ are
\begin{linenomath}
 \begin{equation}
  m_{\mathrm{max}}(s,t) = \frac{16 \Pe_{t}^{2}\mathcal{M}_{0}(s,0^{+},t) \chi(s)^{2}}{3\psi(s)^{2}\mbox{e}^{2}}, \quad n_{\mathrm{max}}(s,t) = \frac{\psi(s)^{2}}{4\Pe_{t}^{2}\chi(s)^{2}},
  \label{eqn:Peak_and_Loc_Diff}
 \end{equation}
\end{linenomath}
with the full-width at half-maximum given by
\begin{linenomath}
 \begin{equation}
  w_{1/2}(s,t) = n_{\mathrm{max}}(s,t)\left[W_{-1}\left(\frac{-1}{\sqrt{2}\mbox{e}}\right)^{2} - W_{0}\left(\frac{-1}{\sqrt{2}\mbox{e}}\right)^{2}\right],
  \label{eqn:whalf_Diff}
 \end{equation}
\end{linenomath}
so that, again, the ring is thicker the further the peak location is from the contact line.

Note that the similarity form and the properties of the nascent coffee ring for both kinetic and diffusive evaporation depend strongly on the behaviour of the local droplet contact angle through $\psi(s)$ and the accumulated mass flux into the boundary $\mathcal{M}_{0}(s,0^{+},t)$. In the diffusive regime, the heterogeneity of evaporation, $\chi(s)$, also plays a role. In \textsection \ref{sec:Elliptical}, we discuss to what extent each factor is relevant in determining the shape of the coffee ring for a specific example. We will also utilize the asymptotic results to discuss the limitations of the dilute assumption, in particular investigating the role that droplet geometry has on the breakdown of the model. 

\subsection{Validation for an axisymmetric droplet} \label{sec:AxisymmetricValidation}

We now seek to validate our results by comparing to the simpler axisymmetric case in which $\Omega$ is simply circular, given by $r = \sqrt{x^{2}+y^{2}} \leq 1$. This case was dealt with in detail by \citet{Moore2021} by proceeding to higher order in the inner region rather than by introducing an intermediate region (as in \textsection \ref{sec:Intermediate}), so that the resulting composite expansions (in \textsection \ref{sec:Composite} and \textsection \ref{sec:Diffusive}) are different. In this section, we validate them by showing excellent agreement with the numerical simulations of \citet{Moore2021}. 

\subsubsection{Kinetic regime}

The simplified geometry allows us to evaluate the leading-order flow solution explicitly. In the kinetic evaporative regime, (\ref{eqn:SmallCaSol1})--(\ref{eqn:PressurePert}) give
\begin{linenomath}
 \begin{equation}
 H = \frac{1-r^{2}}{4}, \; P = P_{0} - \frac{48}{1-r^{2}}, \; \bar{u}_{r} = \frac{\pi r}{4(1-t)}, \; t_{f} = \frac{1}{\pi}, \; \alpha = \frac{8}{\pi}.
 \label{eqn:AxisymmetricSoln}
 \end{equation}
\end{linenomath}
Here $P_{0}$ is an arbitrary constant and $\bar{u}_{r}$ is the radial velocity. (We note that there is a slight difference between (\ref{eqn:AxisymmetricSoln}) and the corresponding results in \citet{Moore2021}: this is due to the fact that the droplet aspect ratio $\delta$ in that paper is defined so that it contains an additional $2/\pi$.)

Now, noting that, for an axisymmetric droplet, $s = \theta$, $n = 1-r$, $\kappa = -1$ and $\bar{u}_{n} = -\bar{u}_{r}$, we can use (\ref{eqn:AxisymmetricSoln}) to determine the leading-order solute concentration in each region, finding from (\ref{eqn:SoluteOuter}), (\ref{eqn:SoluteInner}) and (\ref{eqn:SoluteIntermediate}),
\begin{linenomath}
 \begin{alignat}{2}
  \phi_{0} & \, = && \, \frac{1}{\sqrt{1-t}}\left(\frac{1-\sqrt{1-t}r^{2}}{1-r^{2}}\right), \\
  \hat{\phi}_{0} & \,  = && \, \frac{\pi^{2}}{64(1-t)^{3}}\left[1-\sqrt{1-t}-\frac{t}{2}\right]\exp\left(\frac{-\pi(1-r)}{4\ve(1-t)}\right), \\
  \skn{\phi}_{0} & \, = && \, \mbox{exp}\left(\frac{-\pi(1-r)}{4\ve(1-t)}\right)\left[\frac{\pi^{2}}{64\ve}\frac{(1-\sqrt{1-t}-t/2)}{(1-t)^{3}} + \frac{\pi}{8}\frac{(1-\sqrt{1-t})}{(1-t)^{3/2}}\mbox{Ei}\left(\frac{\pi(1-r)}{4\ve(1-t)}\right)\right].
 \end{alignat}
\end{linenomath}
Combining these expressions and evaluating the overlap contributions, we find that an additive composite expansion for the solute concentration is given by
\begin{linenomath}
 \begin{alignat}{2}
  \phi_{\mathrm{comp}} & \, = && \, \phi_{0}(r,t) + \frac{1}{\ve}\skn{\phi}_{0}(r,t) + \frac{1}{\ve^{2}}\hat{\phi}_{0}(r,t) - \frac{(1-\sqrt{1-t})}{2\sqrt{1-t}}\frac{1}{1-r} \nonumber \\
  & \, && \, -\frac{1}{\ve^{2}}\frac{\pi^{2}(1-\sqrt{1-t}-t/2)}{64(1-t)^{3}}\mbox{exp}\left(-\frac{\pi(1-r)}{4\ve(1-t)}\right) \nonumber \\
  & \, && \, -\frac{1}{\ve}\frac{\pi}{8}\frac{(1-\sqrt{1-t})}{(1-t)^{3/2}}\mbox{exp}\left(-\frac{\pi(1-r)}{4\ve(1-t)}\right)\log\left(\frac{\pi(1-r)}{4\varepsilon(1-t)}\right). 
  \label{eqn:SoluteComposite}
 \end{alignat}
\end{linenomath}
It is worth noting that the composite solution presented here is only valid to $O(1/\ve^{2})$ in the inner region, while the composite solution derived by \citet{Moore2021} is valid to $O(1)$. 

\begin{figure}
\begin{subfigure}
\centering \scalebox{0.4}{\epsfig{file=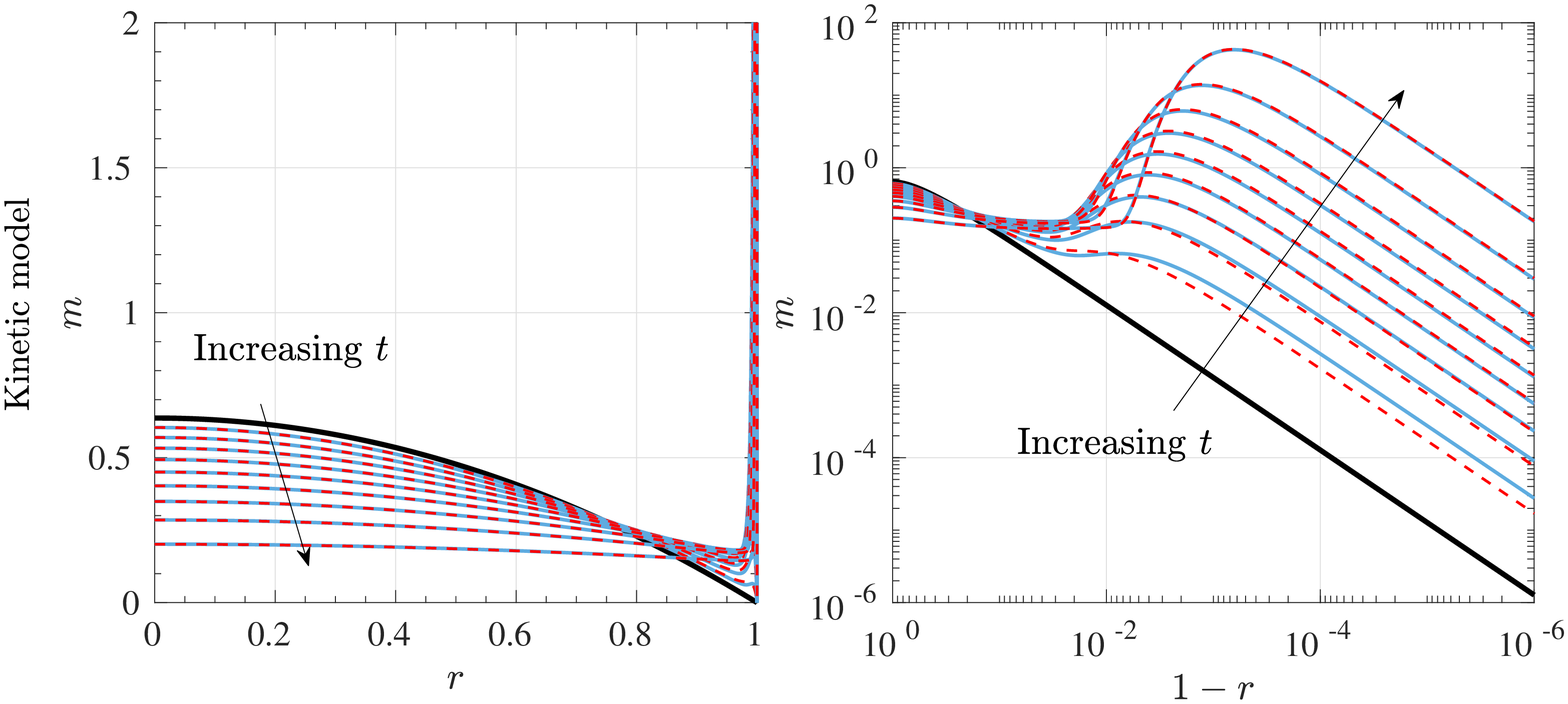}}
\captionsetup{labelformat=empty}
\end{subfigure}
\begin{subfigure}
\centering \scalebox{0.4}{\epsfig{file=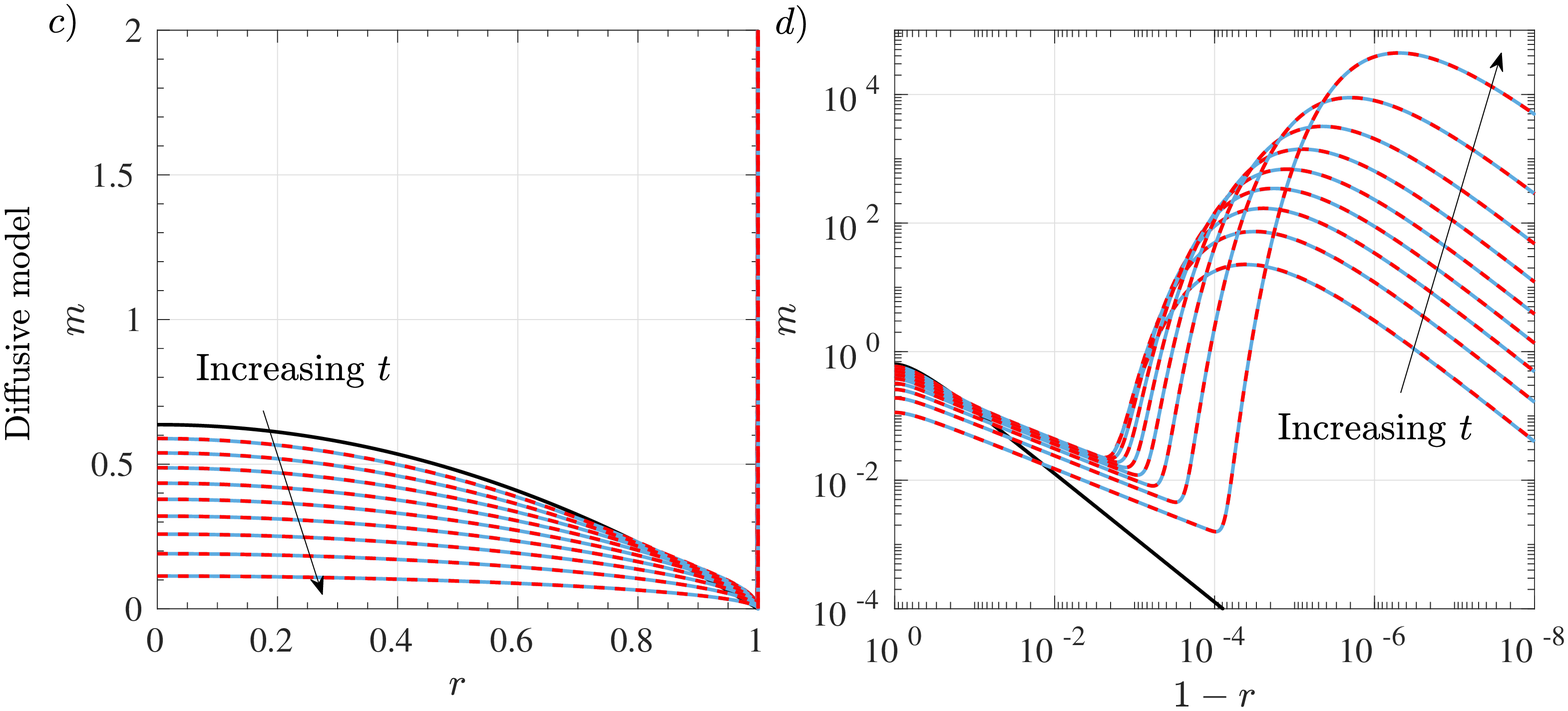}}
\end{subfigure}
\caption{Profiles of the solute mass as an axisymmetric droplet evaporates under (a,b) a kinetic evaporative flux and (c,d) a diffusive evaporative flux with $\Pe = 200$. In each figure, the bold, black curve represents the initial mass profile $m(r) = 2(1-r^{2})/\pi$. Also shown are plots at time intervals of 0.1 up to $t = 0.9$ in which solid, blue curves represent the results from the numerical solution of (\ref{eqn:SoluteNonDim1})--(\ref{eqn:SoluteNonDim3}) and the dashed, red curves show the composite mass profiles, $m = h\phi_{\mathrm{comp}}$. The right-hand figures display a doubly-logarithmic plot of the mass profile near the contact line.}
\label{fig:Validation} 
\end{figure}

\subsubsection{Diffusive regime}

Similarly, in the diffusive regime, (\ref{eqn:SmallCaSol1})--(\ref{eqn:PressurePert}) give
\begin{linenomath}
 \begin{alignat}{2}
 & \, && \, H = \frac{1-r^{2}}{4}, \; \bar{u}_{r} = \frac{1}{r(1-t)}\left(\frac{1}{\sqrt{1-r^{2}}} - (1-r^{2})\right), \; t_{f} = \frac{1}{4}, \; \alpha = \frac{8}{\pi} \nonumber \\
 & \, && \, P = P_{0} - \frac{384}{\pi}\left[\frac{1}{\sqrt{1-r^{2}}} + \frac{1}{3(1-r^{2})^{3/2}} - \frac{1}{2}\log\left(\frac{1-\sqrt{1-r^{2}}}{1+\sqrt{1-r^{2}}}\right) + \log\left(\frac{r}{\sqrt{1-r^{2}}}\right)\right].
 \label{eqn:AxisymmetricSoln_Diffusive}
 \end{alignat}
\end{linenomath}
where, again, $P_{0}$ is an arbitrary constant. The leading-order solute concentration in the outer, inner and intermediate regions are then given by
\begin{linenomath}
 \begin{alignat}{2}
  \phi_{0} & \, = && \, \frac{1}{(1-t)^{1/4}\sqrt{1-r^{2}}}\left[1-(1-t)^{3/4}(1-(1-r^{2})^{3/2})\right]^{1/3}, \\
  \hat{\phi}_{0} & \,  = && \, \frac{1}{24(1-t)^{5}}\left(1-(1-t)^{3/4}\right)^{4/3}\exp\left(\frac{-\sqrt{2(1-r)}}{\ve(1-t)}\right), \\
  \skn{\phi}_{0} & \, = && \, \mbox{exp}\left(-\frac{\sqrt{2(1-r)}}{\ve(1-t)}\right)\left[\frac{1}{24\ve^{3}(1-t)^{5}}\left(1-(1-t)^{3/4}\right)^{4/3} \right. \nonumber \\ 
  & \, && \, \left. + \frac{1}{(1-t)^{5/4}}\left(1-(1-t)^{3/4}\right)^{1/3}\mbox{Ei}\left(\frac{\sqrt{2(1-r)}}{\ve(1-t)}\right)\right].
 \end{alignat}
\end{linenomath}
We can use these to form an additive composite expansion, given by
\begin{linenomath}
 \begin{alignat}{2}
  \phi_{\mathrm{comp}} & \, = && \, \phi_{0}(r,t) + \frac{1}{\ve}\skn{\phi}_{0}(r,t) + \frac{1}{\ve^{4}}\hat{\phi}_{0}(r,t) - \frac{1}{\sqrt{2}(1-t)^{1/4}\sqrt{1-r}}\left(1-(1-t)^{3/4}\right)^{1/3} \nonumber \\
  & \, && \, -\frac{1}{\ve^{4}}\frac{1}{24(1-t)^{5}}\left(1-(1-t)^{3/4}\right)^{4/3}\exp\left(\frac{-\sqrt{2(1-r)}}{\ve(1-t)}\right) \nonumber \\
  & \, && \, -\frac{1}{\ve}\frac{1}{(1-t)^{5/4}}\left(1-(1-t)^{3/4}\right)^{1/3}\mbox{exp}\left(-\frac{\sqrt{2(1-r)}}{\ve(1-t)}\right)\log\left(\frac{\sqrt{2(1-r)}}{\ve(1-t)}\right). 
  \label{eqn:SoluteComposite_Diffusive}
 \end{alignat}
\end{linenomath}

\subsubsection{Comparisons to numerical results}

To check the validity of our asymptotic analysis, we compare profiles of the solute mass $m = h\phi_{\mathrm{comp}}$ against a numerical solution of (\ref{eqn:SoluteNonDim1})--(\ref{eqn:SoluteNonDim3}). While the axisymmetry greatly facilitates the numerical solution, the thinness of the boundary layer as discussed in \textsection \ref{sec:Inner} means that care has to be taken with resolution close to the contact line: we use the numerical scheme validate in \citet{Moore2021}. 

The results are shown for $\Pe = 200$ in figure \ref{fig:Validation}. In the figure, one can clearly see the transport of the solute mass from the droplet bulk towards the contact line as the droplet evaporates, leading to ring formation in the boundary layer. Moreover, it is evident that the asymptotics do an excellent job of capturing the dynamics, particularly as $t$ increases. This gives us confidence in using our asymptotic results to consider the nascent coffee ring characteristics for more complicated geometries, where numerical solutions of (\ref{eqn:SoluteNonDim1})--(\ref{eqn:SoluteNonDim3}) is much more computationally challenging \cite[see, for example,][]{Saenz2017}. 


\section{Droplets with an elliptical contact set} \label{sec:Elliptical}

For the rest of this paper, we shall specialize to droplets that have an elliptical contact set, namely those given by
\begin{linenomath}
 \begin{equation}
  \Omega = \left\{(x,y)\left|\left(\frac{x}{1+a}\right)^{2} + y^{2} \leq 1\right.\right\}
 \end{equation}
\end{linenomath}
where $a \geq 0$ is a constant that encodes the eccentricity, $e(a) = [a(2+a)]^{1/2}/(1+a)$, of the ellipse. 

The forthcoming analysis is more readily approached by introducing the planar elliptical coordinate system $(\mu,\nu)$, which is defined by
\begin{linenomath}
\begin{equation}
 x = \sqrt{2a+a^{2}}\cosh\mu\cos\nu, \; y = \sqrt{2a+a^{2}}\sinh\mu\sin\nu, 
\label{eqn:EllipticalPolars}
\end{equation}
\end{linenomath}
where $0 \leq \mu \leq \mu^{*} = \mbox{acosh}(1/e(a))$ and $\mu = \mu^{*}$ represents the contact line. We use the symmetry of the problem to restrict our analysis to the quarter of the ellipse in the first quadrant for which the fluid domain transforms to the rectangle $0\leq\mu \leq \mu^{*}$, $0\leq\nu\leq\pi/2$.

We note that, in terms of $(\mu,\nu)$, the local coordinate system defined in \textsection \ref{sec:Coordinates} is given by
\begin{linenomath}
\begin{equation}
\begin{array}{r@{}l@{}l}
 s(\nu) & \; = & \; E\left(\nu,\mbox{i}\sqrt{2a+a^{2}}\right), \\[2mm]
 n(\mu,\nu)^{2} & \; = & \; (2a+a^{2})\left[(\cosh\mu^{*}-\cosh\mu)^{2}\cos^{2}\nu 
+(\sinh\mu^{*}-\sinh\mu)^{2}\sin^{2}\nu\right].
\end{array}
\label{eqn:sandnu}
\end{equation}
\end{linenomath}
where $E(\phi,k) = \int_{0}^{\phi} (1-k^{2}\sin^{2}\theta)^{1/2}\,\mbox{d}\theta$ is the incomplete elliptical integral of the second kind with amplitude $\phi$ and elliptic modulus $k$. 

Unlike the case of an axisymmetric drop, the $(s,n)$-coordinate system is not well-defined throughout the whole quarter-ellipse. For the system to be well-defined at a point, we require a unique normal through that point, and hence a unique value of $(s,n)$. This is true everywhere in the quarter ellipse aside from the interval $0 \leq x \leq 1-1/(1+a)$ along the major semi-axis, so that we need $1/(1+a) \gg 1/\Pe$ (or $1/\Pe^{2}$ for the diffusive regime) for our analysis in \textsection \ref{sec:Inner} to be valid. This is satisfied (for $\Pe\gg1$) provided that $a\ll\Pe$, so our analysis is limited to eccentricities $e(a) \lesssim 1-1/2\Pe^{2}$. We also note that since the relation between $s$ and $\nu$ in (\ref{eqn:sandnu}) is independent of $\mu$, we can use these interchangeably in the rest of our analysis; for convenience, we shall use $\nu$. It is also important to note that both (\ref{eqn:EllipticalPolars}) and (\ref{eqn:sandnu}) are \emph{contact set dependent}, in that they change with $a$. In particular, while $\nu$ heuristically indicates the angular position of a particular point, the $\nu$-coordinate of a point with one ellipse eccentricity is not the same for an ellipse with a different eccentricity; the exceptions to this are the points on the semi-major and semi-minor axes, for which $\nu = 0$ and $\nu = \pi/2$ for all $a$. Hence, in the following, when we wish to make explicit comparisons between different ellipses, we shall focus on the semi-axes; happily, this is also where the effects of contact line curvature are seen most clearly.

Our aim in this section is to illustrate the effect of the droplet geometry on the nascent coffee ring. To affect sensible comparisons for elliptical contact sets with different eccentricities, we will consider droplets that have the same initial volume, $V^{*}$, and contact line perimeter, $P^{*}$. When changing the droplet shape, but keeping $V^{*}$ and $P^{*}$ fixed, we change the characteristic length scaling, $R^{*}$, and velocity scaling, $U^{*}$, in our model, as well as the values of $\delta$ and $\Pe$. Here we discuss how these change when comparing an axisymmetric droplet to an elliptical droplet of the same volume and perimeter.

For an axisymmetric droplet ($e(0) = 0$), we denote these quantities by a subscript zero. Thus, taking the characteristic lengthscale to be the radius of the circular constant set, we have
\begin{linenomath}
 \begin{equation}
  R_{0} = \frac{P^{*}}{2\pi}, \; \delta = \frac{V^{*}}{R_{0}^{*3}}, \; U^{*}_{0} = \frac{\mathcal{E}^{*}}{\rho\delta_{0}}, \; \Pe_{0} = \frac{R_{0}^{*}U_{0}^{*}}{D^{*}}.
 \end{equation}
\end{linenomath}

For an elliptical droplet, recall that we took the dimensional length of the semi-minor axis as our reference lengthscale for the size of the contact set, $R^{*}$. This lengthscale changes with the eccentricity of the droplet as encoded through $a$. Therefore, using a subscript $a$ to denote the different properties, the equivalent values for an ellipse are given by
\begin{linenomath}
\begin{equation}
\begin{aligned}
  R_{a}^{*} & \, = 2\pi \left(\int_{0}^{2\pi}\left[(1+a)^{2}\cos^{2}\theta + \sin^{2}\theta\right]\,\mbox{d}\theta\right)^{-1}R_{0}^{*}, \\
  & \, \delta_{a} = \frac{\delta_{0}R_{0}^{*3}}{R_{a}^{*3}}, \; U_{a}^{*} = \frac{\delta_{0}U_{0}^{*}}{\delta_{a}}, \; \Pe_{a} = \frac{R_{a}^{*}U_{a}^{*}}{R_{0}^{*}U_{0}^{*}}\Pe_{0}.
  \label{eqn:R_and_d_ratios}
\end{aligned}
\end{equation}
\end{linenomath}

\subsection{Free surface profile} \label{sec:FreeSurface_Ellipse}

In an elliptical geometry, the Poisson problem (\ref{eqn:Poisson}) can be solved explicitly, yielding
\begin{linenomath}
 \begin{equation}
  H = \frac{1}{2}\left(1+\frac{1}{(1+a)^{2}}\right)^{-1}\left[1-\frac{2a+a^{2}}{(1+a)^{2}}\cosh^{2}\mu\cos^{2}\nu - (2a+a^{2})\sinh^{2}\mu\sin^{2}\nu\right]. 
  \label{eqn:Poisson_sol_ellipse}
 \end{equation}
\end{linenomath}
We can use this free surface profile to determine the constant $\alpha$ used to rescale volume, see (\ref{eqn:alpha}), finding
\begin{linenomath}
 \begin{equation}
  \alpha = \frac{4}{\pi(1+a)}\left(1+\frac{1}{(1+a)^{2}}\right).
 \end{equation}
\end{linenomath}
It is worth noting that if we expand (\ref{eqn:Poisson_sol_ellipse}) as the contact line is approached, $\mu^{*}-\mu\rightarrow0$, we have 
\begin{linenomath}
 \begin{equation}
  H \sim \frac{1}{(1+a)}\left(1+\frac{1}{(1+a)^{2}}\right)^{-1}\left(1 + (2a+a^{2})\sin^{2}\nu\right)(\mu^{*}-\mu), \label{eqn:MuNuAsymp1}
  \end{equation}
\end{linenomath}
so that, recalling (\ref{eqn:sandnu}) to determine $\mu^{*}-\mu$ as a function of $n$, the rescaled contact angle, $\psi = \theta_{c}(\nu,t)/(1-t)$, may be found from (\ref{eqn:ContactAngle}), (\ref{eqn:sandnu}) and (\ref{eqn:MuNuAsymp1}) to be
\begin{linenomath}
 \begin{equation}
   \psi(\nu) = \frac{4}{\pi(1+a)^{2}}\left[1 + (2a+a^{2})\sin^{2}\nu\right]^{1/2}. \label{eqn:EllipticA}
 \end{equation}
\end{linenomath}

\subsection{Diffusive evaporative flux and dryout times} \label{sec:DiffusiveFlux_Ellipse}

As described in \citet{Kellogg1929}, it is possible to solve the concentration problem (\ref{eqn:DiffusionVapour}) for an elliptical contact set. The resulting evaporative flux is given by
\begin{linenomath}
 \begin{equation}
  E(\mu,\nu) = \frac{1}{K[e(a)]}\left[1-\frac{2a+a^{2}}{(1+a)^{2}}\cosh^{2}\mu\cos^{2}\nu - (2a+a^{2})\sinh^{2}\mu\sin^{2}\nu\right]^{-1/2}
  \label{eqn:DiffusiveFlux_Ellipse}
 \end{equation}
\end{linenomath}
where $K(k) = \int_{0}^{\pi/2}(1-k^{2}\sin^{2}\theta)^{-1/2}\,\mbox{d}\theta$ is the complete elliptic integral of the first kind with elliptic modulus $k$. Expanding (\ref{eqn:DiffusiveFlux_Ellipse}) close to the contact line, we have $E \sim \chi(\nu)n^{-1/2}$ with
\begin{linenomath}
 \begin{equation}
  \chi(\nu) = \frac{1}{K[e(a)]}\left(\frac{1+a}{2}\right)^{1/2}\left(1 + (2a+a^{2})\sin^{2}\nu\right)^{-1/4}.
  \label{eqn:DiffVapLocal_ellipse}
 \end{equation}
\end{linenomath}

The dryout time is evaluated from (\ref{eqn:EvaporationTimeNonDim}) yielding
\begin{linenomath}
 \begin{equation}
  t_{f} = \begin{cases}
           \displaystyle{\frac{1}{\pi(1+a)}} & \mbox{in the kinetic regime,} \\[3mm]
           \displaystyle{\frac{K[e(a)]}{2\pi(1+a)}} & \mbox{in the diffusive regime.}
          \end{cases}
 \end{equation}
\end{linenomath}

\subsection{Fluid velocity}

Unfortunately, no such analytical progress is possible for the pressure problem (\ref{eqn:PressurePert}), which must be solved numerically for each evaporative model. We have found that a convenient way to approach this is to subtract out the most singular terms in $P$ at the contact line; we describe this process and the details of our numerical methodology in Appendix \ref{appendix:Numerics}. 

We can, however, make some comments about the velocity close to the contact line, making use of the fact that in the local coordinate system, the normal velocity is given by
\begin{linenomath}
\begin{equation}
 \bar{u}_{n} = -\bar{u}_{\mu}.
 \label{eqn:unandumu}
\end{equation}
\end{linenomath}

\subsubsection{Kinetic evaporation}

\begin{figure}
\centering \scalebox{0.425}{\epsfig{file=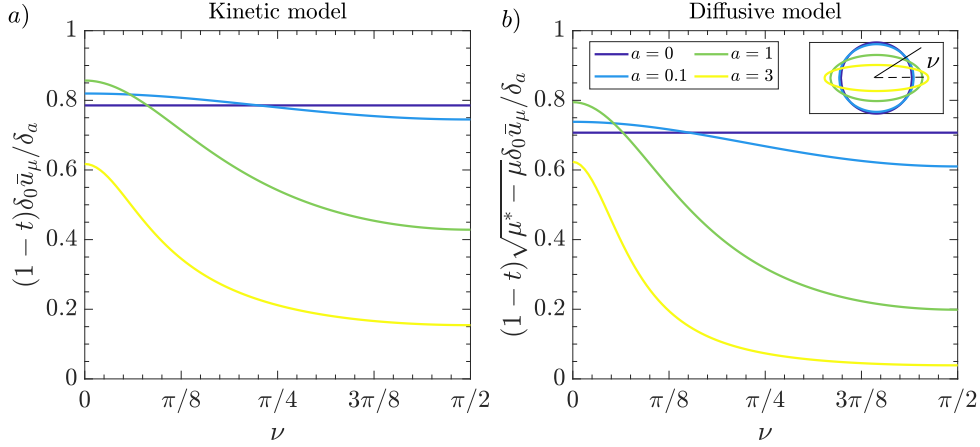}}
 \caption{The angular dependence of the normal velocity close to the contact line for $a)$ a kinetic evaporative model as given by (\ref{eqn:MuNuAsymp2}), and $b)$ a diffusive evaporative model as given by (\ref{eqn:MuNuAsymp3}). In each figure, we vary the shape of the elliptical contact set by changing $a$, but scale in such a way that each droplet has the same initial volume and perimeter. For illustration, we show the different droplet footprints in the inset to figure $b)$. Displayed are the axisymmetric case, $a = 0$ ($e(0) = 0$, dark purple), $a = 0.1$ ($e(0.1) \approx 0.417$, blue), $a = 1$ ($e(1) \approx 0.866$, green) and $a = 3$ ($e(3) \approx 0.968$, yellow).} 
 \label{fig:KineticVelocityProfile}
 \end{figure}

For the kinetic evaporative model, we combine (\ref{eqn:Un_local}) and (\ref{eqn:EllipticA}) to show that
\begin{linenomath}
 \begin{equation}
  \bar{u}_{\mu} \sim \frac{\pi(1+a)^{2}}{4(1-t)}\left[1+(2a+a^{2})\sin^{2}\nu\right]^{-1/2}, \label{eqn:MuNuAsymp2}
  \end{equation}
\end{linenomath}
as $\mu^{*}-\mu\rightarrow0$. To illustrate the effects of the droplet geometry on the velocity profile, we display the normal velocity close to the contact line (\ref{eqn:MuNuAsymp2}) as a function of the elliptical polar angle $\nu$ in figure \ref{fig:KineticVelocityProfile}a for different eccentricities. We have scaled $(1-t)\bar{u}_{\mu}$ by $\delta_{0}/\delta_{a}$ so that for each curve the droplet has the same initial volume and perimeter, with just the eccentricity of the ellipse changing (cf. (\ref{eqn:R_and_d_ratios})). The axisymmetric case is illustrated by the dark purple line. Initially, as we increase $a$ it is clear that the velocity is increased in the regions with higher curvature, namely close to the semi-major axis, while being diminished close to the minor axis. However, since $\delta_{a}$ increases with $a$, eventually the velocity is lower than the equivalent axisymmetric problem around the whole contact line. Nevertheless, the velocity is still relatively stronger along the more highly-curved parts of the boundary, and this disparity grows as $a$ increases: that is, for more eccentric ellipses, the stronger the flow in the direction of the semi-major axis in comparison to that along the semi-minor axis.

\subsubsection{Diffusive evaporation}

Similarly, for a diffusive evaporative model, combining  (\ref{eqn:Un_local_diffusive}), (\ref{eqn:sandnu}), (\ref{eqn:EllipticA}) and (\ref{eqn:DiffVapLocal_ellipse}) gives
\begin{linenomath}
 \begin{equation}
  \bar{u}_{\mu} \sim \frac{\pi(1+a)^{5/2}}{2^{3/2}K(e(a))(1-t)}\frac{1}{1+(2a+a^2)\sin^{2}\nu}\left(\mu^{*}-\mu\right)^{-1/2}
  \label{eqn:MuNuAsymp3}
 \end{equation}
\end{linenomath} 
as $\mu^{*}-\mu\rightarrow0$. We plot $(1-t)\sqrt{\mu^{*}-\mu}\delta_{0}\bar{u}_{\mu}/\delta_{a}$ for different ellipse eccentricities in figure \ref{fig:KineticVelocityProfile}b, where we again scale appropriately to fix the initial droplet volume and perimeter. We see very similar behaviour to the kinetic regime: for all ellipse eccentricities, the normal velocity is stronger along the semi-major axis than the semi-minor axis, with this effect being amplified as $a$ increases. For $a$ small, the velocity is also stronger along the semi-major axis compared to the equivalent axisymmetric droplet, but as $a$ gets larger, the velocity is weaker everywhere around the contact line.

\subsection{Accumulated mass flux}

\subsubsection{Kinetic evaporation}

We can use the local expansions for the outer solute concentration, (\ref{eqn:Phi_local}), the contact angle, (\ref{eqn:MuNuAsymp1}), and the velocity, (\ref{eqn:MuNuAsymp2}), together with (\ref{eqn:sandnu})--(\ref{eqn:unandumu}) to express the accumulated mass flux into the contact line for a kinetic evaporative model as
\begin{linenomath}
 \begin{equation}
  \mathcal{M}_{0}(\nu,0,t) = \frac{1}{\pi(1+a)}\left[1+(2a+a^{2})\sin^{2}\nu\right]^{1/2}\int_{0}^{t}B(\nu,\tau)\,\mbox{d}\tau. \label{eqn:EllipticMathcalM}
 \end{equation}
\end{linenomath}

To investigate how $\mathcal{M}_{0}$ varies with $\nu$ and $a$, however, we need to determine $B(\nu,t)$, which must be found numerically. The solution procedure is therefore as follows. Firstly, we solve (\ref{eqn:PressurePert}) numerically to determine $\bar{\mbf{u}}$ from (\ref{eqn:SmallCaSol3}). Since (\ref{eqn:PressurePert}) is independent of $t$, we need only do this once. We can then use the velocity profile to solve the leading-order-outer solute problem, (\ref{eqn:SoluteOuter}) --- again numerically --- which then allows us to determine $B(\nu,t)$. Finally, we can use $B(\nu,t)$ to evaluate the integral in (\ref{eqn:EllipticMathcalM}). The details of the numerical methodologies for each step are recorded in Appendix \ref{appendix:Numerics}. 

\begin{figure}
\begin{subfigure}
\centering \scalebox{0.42}{\epsfig{file=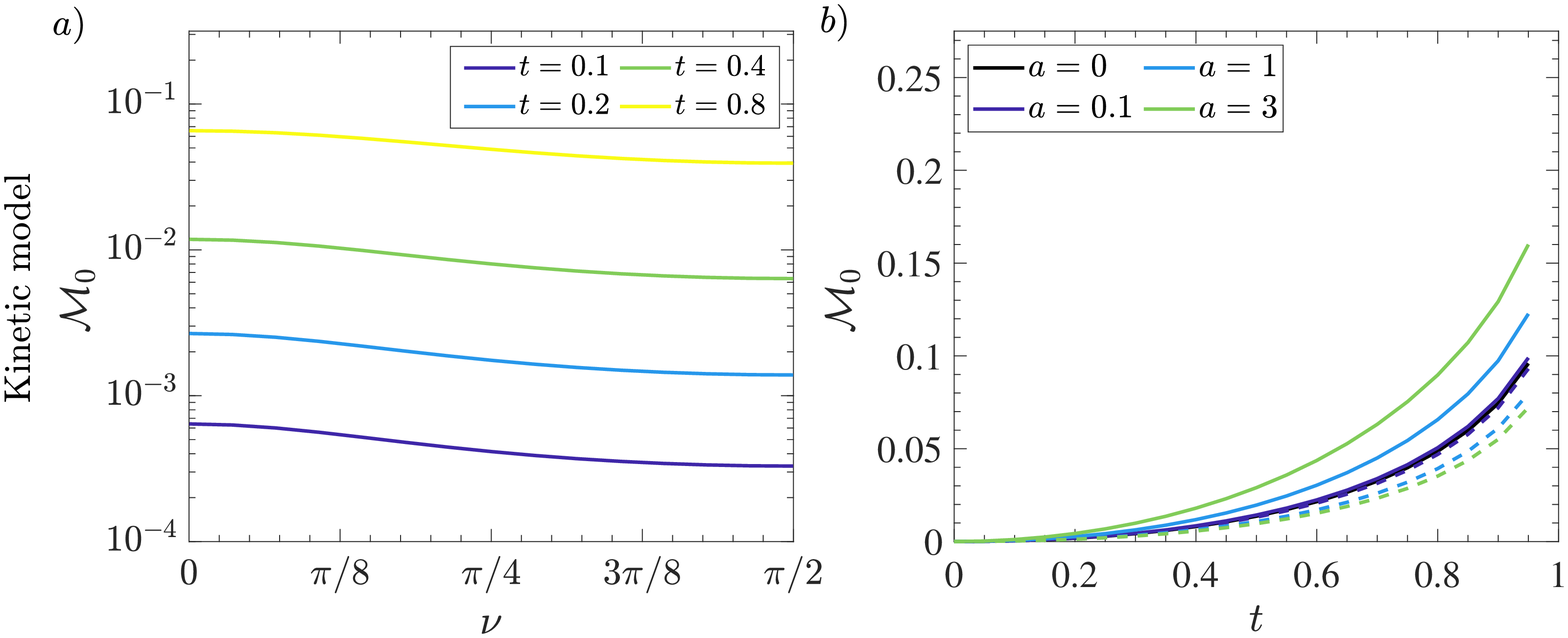}}
\captionsetup{labelformat=empty}
\end{subfigure}
\begin{subfigure}
 \centering \scalebox{0.42}{\epsfig{file=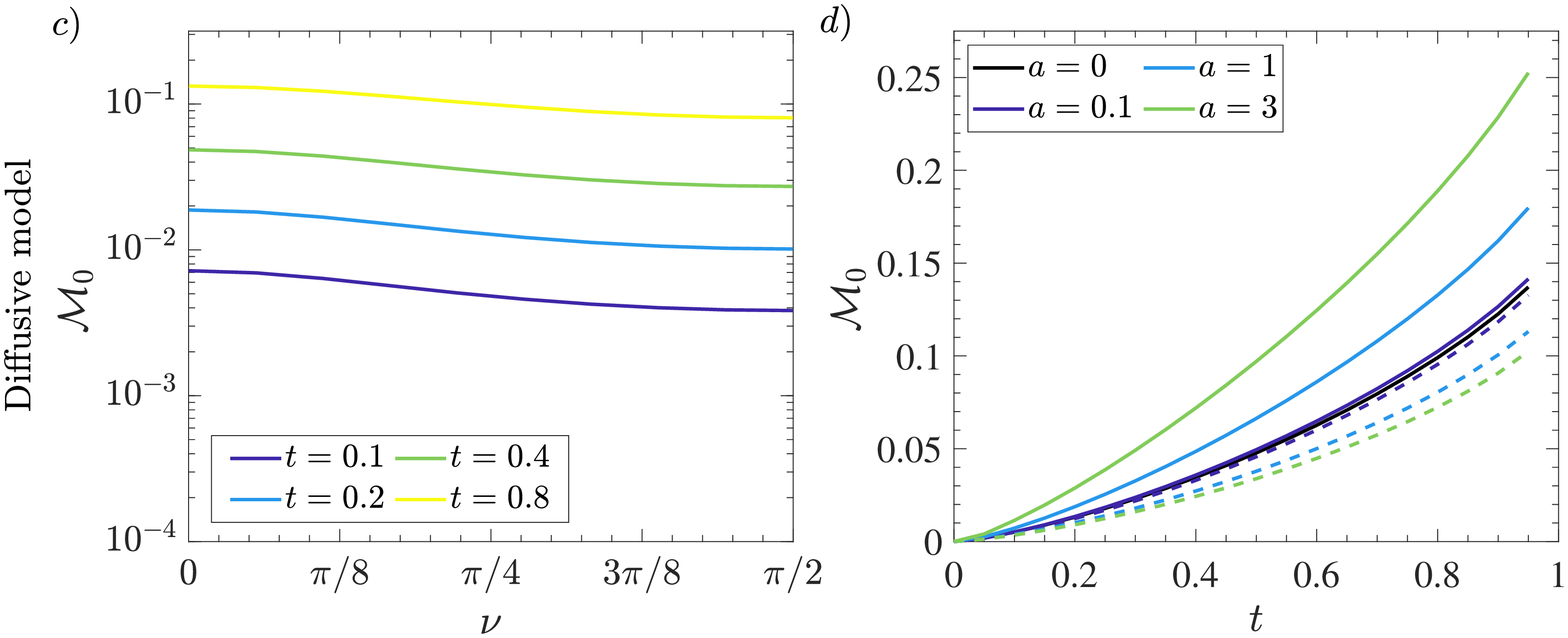}}
 \end{subfigure}
\caption{The angular dependence of the accumulated mass flux at the contact line under $(a, b)$ a kinetic evaporative flux, as given by (\ref{eqn:EllipticMathcalM}), and $(c, d)$ a diffusive evaporative flux, as given by (\ref{eqn:EllipticalMathcalM_Diffusive}). $a), c)$ Profiles of $\mathcal{M}_{0}(\nu,0,t)$ at $t = 0.2$ (dark purple), $t = 0.4$ (blue), $t = 0.6$ (green) and $t = 0.8$ (yellow) for an ellipse with $a = 1$ ($e(1) \approx 0.866$). $b), d)$ The accumulated mass flux along the major (solid) and minor (dashed) semi-axes for different ellipse eccentricities: $a = 0$ (black), $a = 0.1$ ($e(0.1) \approx 0.417$, dark purple), $a = 1$ ($e(1) \approx 0.866$, blue) and $a = 3$ ($e(3) \approx 0.968$, green). In each of $(b,d)$, we have scaled appropriately so that the droplets have the same perimeter and initial volume.}
 \label{fig:MassFluxes}
 \end{figure}

We display the resulting accumulated mass flux into the contact line (\ref{eqn:EllipticMathcalM}) in figure \ref{fig:MassFluxes}a,b. In figure \ref{fig:MassFluxes}a, we display $\mathcal{M}_{0}$ as a function of the elliptical polar angle $\nu$ at different stages of the evaporation for $a = 1$. Clearly, even at small times, there is a larger mass flux accumulating along the parts of the contact line with higher curvature, consistent with the results of \citet{FreedBrown2015} and \citet{Saenz2017}. This disparity increases as $t$ increases. We display the two extremes by plotting $\mathcal{M}_{0}$ along each semi-axis for various values of the eccentricity $a$ in figure \ref{fig:MassFluxes}b. As previously, we have used the scalings (\ref{eqn:R_and_d_ratios}) to compare droplets that have the same initial volume and contact line perimeter. It is apparent from the figure that, as $a$ increases, the accumulated mass flux along the semi-major axis increases compared to the axisymmetric case, while the accumulated mass flux along the semi-minor axis decreases compared to the same. These behaviours are accentuated further as the ellipse becomes more eccentric. It is worth stressing that this is in spite of the normal velocity along $\partial\Omega$ being smaller everywhere than in the equivalent axisymmetric problem for large values of $a$ (as seen in figure \ref{fig:KineticVelocityProfile}a). 

\subsubsection{Diffusive evaporation}

The equivalent expression for the accumulated mass flux under a diffusive evaporative model is determined from (\ref{eqn:EllipticA}), (\ref{eqn:MuNuAsymp3}) and (\ref{eqn:EllipticalMathcalM_Diffusive}) to be
\begin{linenomath}
 \begin{equation}
  \mathcal{M}_{0}(\nu,0,t) = \frac{1}{\pi}\sqrt{\frac{1}{2(1+a)}}\left[1+(2a+a^{2})\sin^{2}\nu\right]^{-1/4}\int_{0}^{t}B_{d}(\nu,\tau)\,\mbox{d}\tau. \label{eqn:EllipticalMathcalM_Diffusive}
 \end{equation}
\end{linenomath}
Again, we must determine $B_{d}(\nu,\tau)$ numerically by solving the leading-order-outer solute advection problem for $\phi_{0}$. The procedure is identical to the kinetic regime and we display the resulting profiles of $\mathcal{M}_{0}(\nu,0,t)$ in figure \ref{fig:MassFluxes}c,d. The broad behaviour of accentuated mass accumulation along the semi-major axis is very similar to the kinetic regime, although it is notable that $\mathcal{M}_{0}(\nu,0,t)$ is larger at earlier times in the diffusive regime. While under both evaporative models all of the mass will be driven to the contact line at the dryout time (since diffusion is a lower order effect in the outer region), the stronger evaporative flux in the diffusive regime means that mass accumulates faster at the contact line in this regime. 

\subsection{Mass profiles} \label{sec:MassProfiles}

\begin{figure}
\begin{subfigure}
\centering \scalebox{0.32}{\epsfig{file=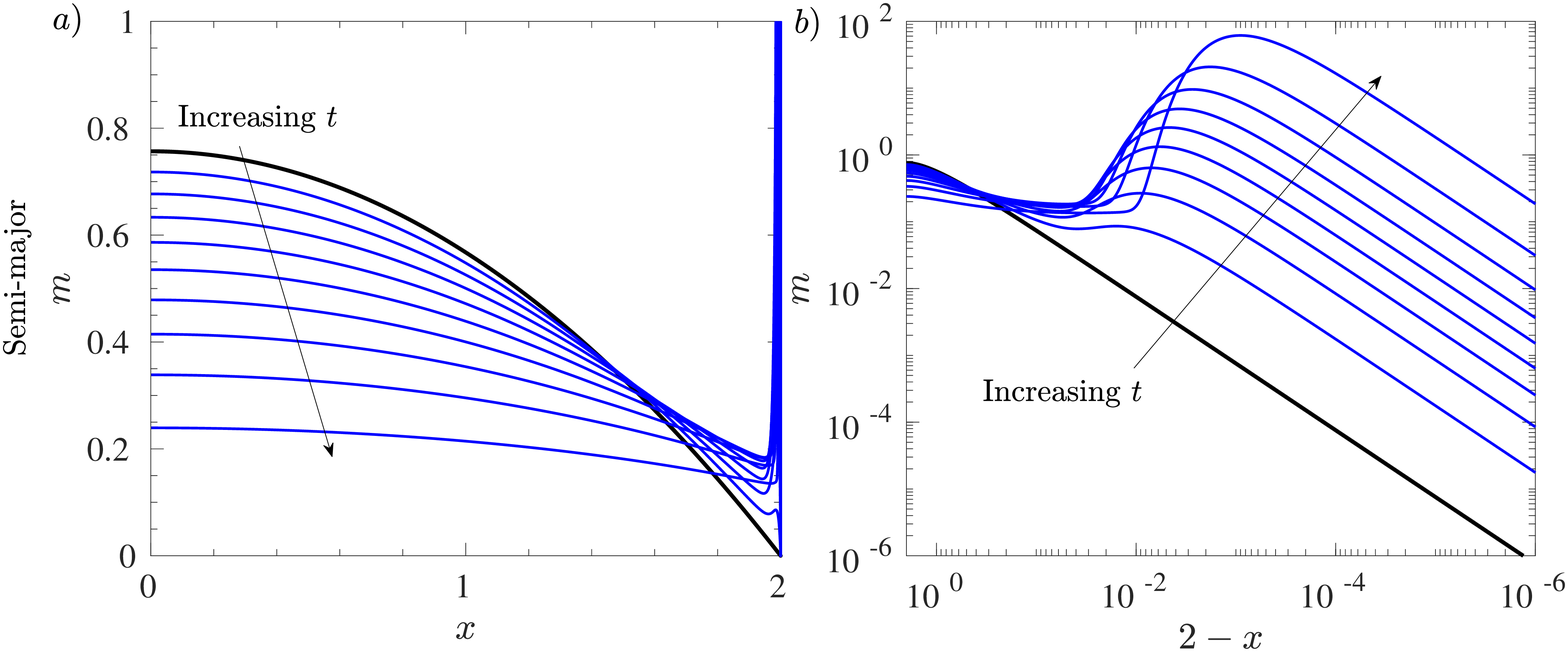}}
\captionsetup{labelformat=empty}
\end{subfigure}
\begin{subfigure}
\centering \scalebox{0.32}{\epsfig{file=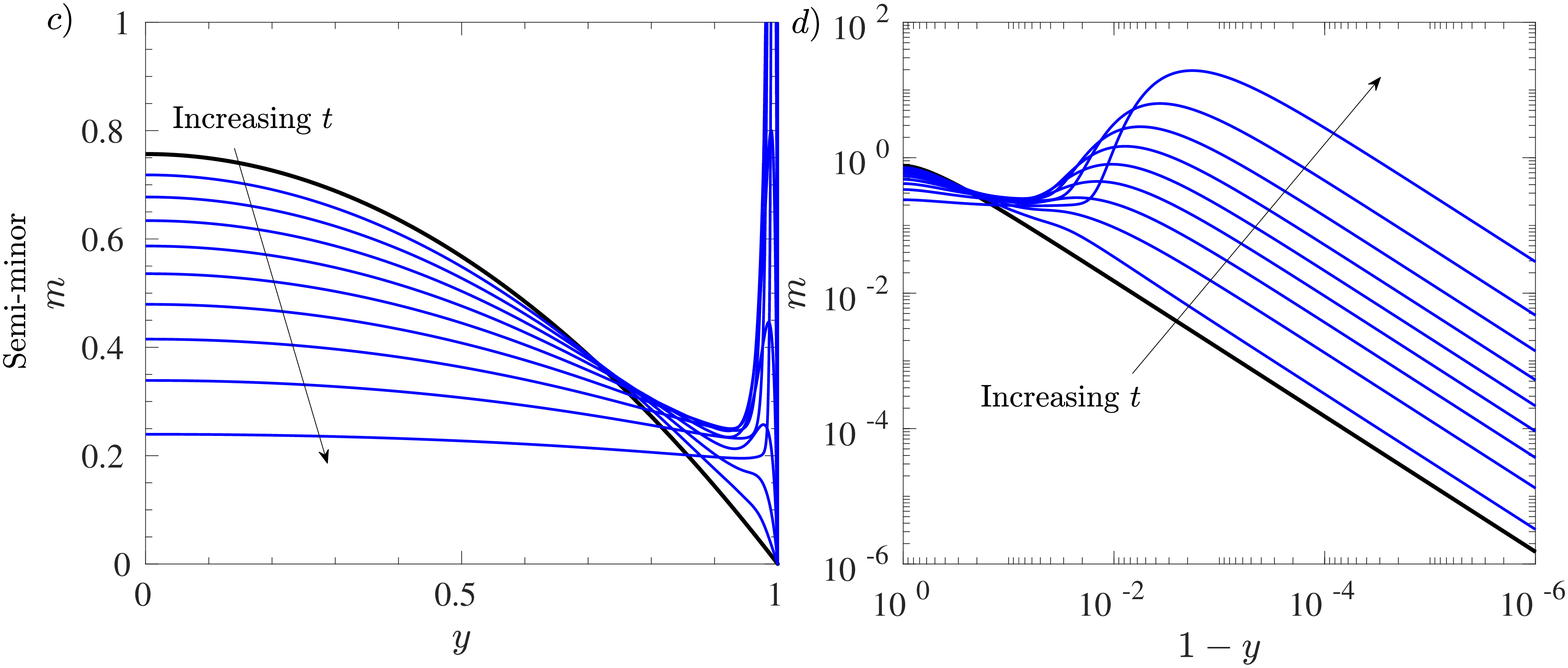}}
\end{subfigure}
\caption{Profiles of the solute mass $m = h\phi_{\mathrm{comp}}$ along the semi-major ($a, b$) and semi-minor ($c, d$) axes of an elliptical droplet with $a = 1$ ($e(1) = 0.866$) evaporating under a kinetic flux for $\Pe \approx 35$ (so that the droplet has the same initial volume and perimeter as the axisymmetric droplet shown in figure \ref{fig:Validation}). In each figure, the bold, black curve represents the initial mass profile, while plots at time intervals of 0.1 up to $t = 0.9$ are shown as solid, blue curves. Figures $b, d$ display a doubly-logarithmic plot of the mass profiles near the contact line, where we see the formation of the nascent coffee ring as $t$ increases. It is notable that the coffee ring effect is much stronger along the semi-major axis due to the strong accumulated mass flux in this region.}
\label{fig:ProfileComparisons} 
\end{figure}

With the accumulated mass flux $\mathcal{M}_{0}$ in hand, we are able to construct the leading-order solute mass profiles, $m = h\phi_{\mathrm{comp}}$, using (\ref{eqn:SoluteCompositeGeneral}) in the kinetic regime and (\ref{eqn:SoluteCompositeDiffusive}) in the diffusive regime. Our aim is to consider the relative influence of the droplet geometry and the evaporative flux on the resulting coffee-ring profile.


\subsubsection{Kinetic evaporation}

To isolate the role of geometry-induced fluid flow, we first consider the kinetic evaporative model. We display in figure \ref{fig:ProfileComparisons} mass profiles along the semi-minor and semi-major axes for an ellipse with $a = 1$ ($e(1) = 0.866$). The droplet has been chosen to have the same initial volume and perimeter as the axisymmetric droplet depicted in figure \ref{fig:Validation}, so that the equivalent P\'{e}clet number is $\Pe_{a} \approx 35$. In the figure, the bold black lines represent the initial mass profile, while the blue lines represent the mass profile evolution along the major axis ($a, b$) and minor axis ($c, d$). As the droplet evaporates, we can clearly see the formation of the nascent coffee ring along both semi-axes, with a characteristic thin, sharp peak growing close to the pinned contact line. It is noticeable that the peak along the semi-major axis is larger than that along the minor. At $90\%$ of the drying time, the coffee ring peak is approximately $3$ times higher along the semi-major axis than the semi-minor axis. The coffee-ring effect is also enhanced when compared to the axisymmetric case: the peak along the semi-major axis is approximately $40\%$ larger than that for the equivalent axisymmetric droplet shown in figure \ref{fig:Validation}, with the peak on the minor axis approximately $2.2$ times smaller.

It is also worth noting that the peak along the semi-minor axis is slightly further from the contact line compared to the semi-major axis. For example (and accounting for the rescalings in (\ref{eqn:R_and_d_ratios})), at $90\%$ of the drying time, the peak along the semi-major axis is at $2-x \approx 5.8\times10^{-4}$, while along the semi-minor axis it is at $1-y \approx 1.3\times10^{-3}$. For reference, the axisymmetric peak location is comparable to the peak on the semi-major axis, with $1-r \approx 6.5\times10^{-4}$ in figure \ref{fig:Validation}.

\begin{figure}
\begin{subfigure}
\centering \scalebox{0.4}{\epsfig{file=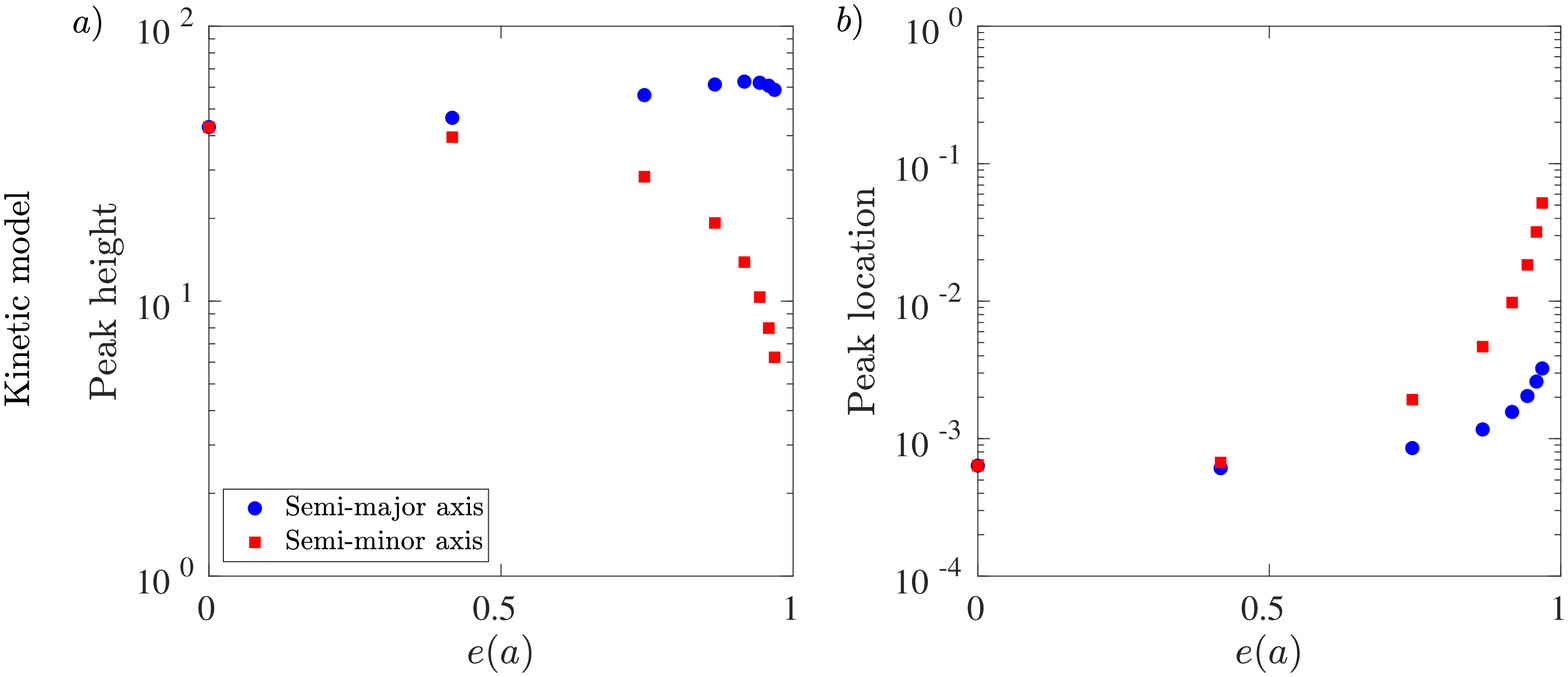}}
\captionsetup{labelformat=empty}
\end{subfigure}
\begin{subfigure}
\centering \scalebox{0.4}{\epsfig{file=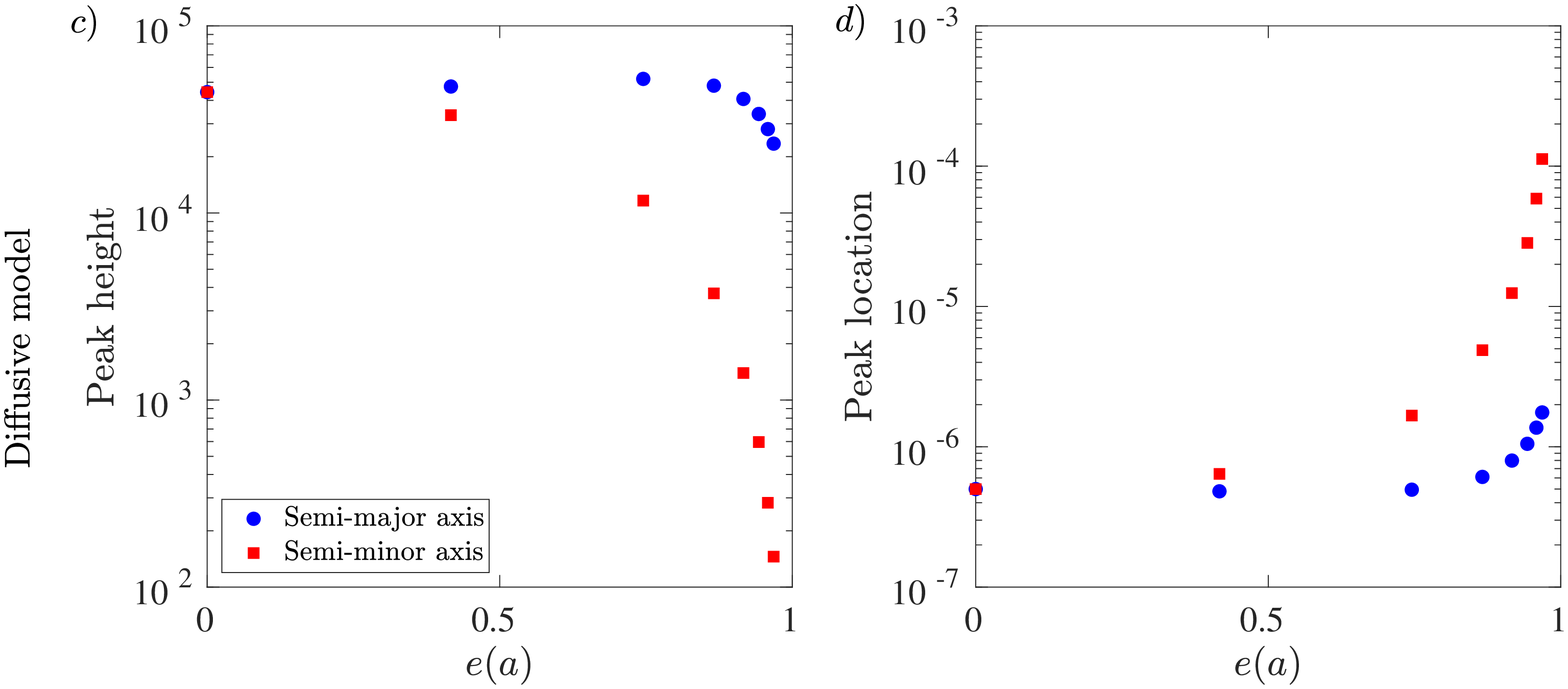}}
\end{subfigure}
\caption{Variation with eccentricity of the dimensionless peak coffee-ring height ($a, c$) and its dimensionless distance from the contact line ($b, d$) at $90\%$ of the drying time. Each ellipse has the same initial perimeter and volume. We display results for both the kinetic ($a, b$) and diffusive ($c, d$) regimes. In each figure, the red squares indicate results for the semi-minor axis, while the blue circles indicate results for the semi-major axis.}
\label{fig:Fixed_V_and_P_kinetic} 
\end{figure}

We further illustrate the effect of ellipse geometry in figure \ref{fig:Fixed_V_and_P_kinetic}a,b, where we show the variation of the maximum coffee-ring peak (a) and its distance from the contact line (b) along each semi-axis with ellipse eccentricity. Each ellipse has the same initial volume and perimeter, with the P\'{e}clet number of the axisymmetric drop taken to be $\Pe_{0} = 200$. The equivalent P\'{e}clet numbers for the elliptical drops are then calculated from (\ref{eqn:R_and_d_ratios}). The results are shown at $90\%$ of the drying time. 

Initially, as the ellipse eccentricity is increased, the peak height along the semi-major axis increases, until reaching a maximum at $a \approx 1.5$ $(e(a) \approx 0.917)$, where it is approximately $1.46$ times larger than the equivalent axisymmetric droplet. For larger eccentricities, the peak height then decreases again, although remaining higher than the axisymmetric case for the eccentricities displayed. It is worth noting that as $e(a)$ approaches unity, the assumptions made in deriving the model begin to break down, with the aspect ratio of the droplet contact becoming larger than $O(1)$. Moreover, the P\'{e}clet number as given by (\ref{eqn:R_and_d_ratios}) decreases with $a$, so the existence of a maximum is not unexpected.

On the other hand, the peak height along the semi-minor axis decreases monotonically as the eccentricity of the ellipse increases. Indeed, as $e(a)$ gets closer to unity, the rate of decrease of the height gets faster. For $a = 3$ ($e(a) \approx 0.968$), the peak height along the semi-minor axis has decreased by almost a factor of $7$. 

Along the semi-minor axis, the location of the coffee ring peak moves radially inwards away from the pinned contact line of the droplet. Correspondingly, according to the analysis of \textsection \ref{sec:Similarity}, as the ellipse eccentricity increases, the thickness of the nascent coffee ring as measured by the full-width at half-maximum increases. Along the semi-major axis, the distance of the peak from the contact line location decreases slightly at very small eccentricities, before again becoming larger than the equivalent axisymmetric case as $e(a)$ approaches unity. The effect is noticeably weaker than that along the semi-minor axis. 

Thus, in summary, on the semi-minor axis, as the eccentricity of the ellipse increases, the coffee ring gets progressively shallower and wider as compared to an axisymmetric droplet of the same initial volume and perimeter. On the semi-major axis, the coffee ring initially becomes narrower and higher, before transitioning to a ring that is wider and higher than the equivalent axisymmetric droplet.

These behaviours can be further probed by considering the similarity analysis of \textsection \ref{sec:Similarity}. The local similarity profile is given by (\ref{eqn:SimilarityForm}), which becomes
\begin{linenomath}
 \begin{equation}
  \frac{\hat{m}_{0}}{\Pe_{t}\mathcal{M}_{0}(\nu,0,t)} = \frac{\pi^{2}(1+a)^{4}}{16}\frac{N}{[1+(2a+a^{2})\sin^{2}\nu]}\mbox{exp}\left(\frac{-\pi(1+a)^{2}N}{4[1+(2a+a^{2})\sin^{2}\nu]^{1/2}}\right), \label{eqn:SimilarityEllipse}
 \end{equation}
\end{linenomath}
where $N = \Pe_{t}n$ and $n$ is given by (\ref{eqn:sandnu}). The leading-order coffee ring peak and its location are found from (\ref{eqn:MaximumAndLocation}) to be
\refstepcounter{equation}
\begin{linenomath}
 $$
  \frac{m_{\mathrm{max}}}{\Pe_{t}\mathcal{M}_{0}(\nu,0,t)} = \frac{\pi(1+a)^{2}}{4\mbox{e}(1+(2a+a^{2})\sin^{2}\nu)^{1/2}}, \; \Pe_{t}n_{\mathrm{max}} = \frac{4(1+(2a+a^{2})\sin^{2}\nu)^{1/2}}{\pi(1+a)^{2}}.
  \eqno{(\theequation{\mathit{a},\mathit{b}})}
  \label{eqn:MaxEllipse}
 $$
\end{linenomath}

It is clear that the right-hand side of (\ref{eqn:MaxEllipse}a) is monotonically decreasing in $\nu$. In particular, its value along $\nu = 0$ is $(1+a)^{2}$ times larger than that along $\nu = \pi/2$. Thus, the geometry-induced flow alone drives an enhanced coffee ring along the semi-major axis. Moreover, we note that since, for a fixed $a$, $\mathcal{M}_{0}(\nu,0,t)$ is a decreasing function of $\nu$ (cf.  figure \ref{fig:KineticVelocityProfile}c), including the accumulated mass flux accentuates this effect further. 

The peak location and hence the full-width at half-maximum (cf. (\ref{eqn:HalfWidth})) are independent of the accumulated mass flux, so that their behaviour as functions of $\nu$ is purely governed by the local free surface profile. As is clearly seen from (\ref{eqn:MaxEllipse}b), as $\nu$ increases, the ring peak is further from the contact line, leading to a thicker coffee ring. The scale factor in terms of distance from the contact line --- and hence thickness of the ring --- between the major and minor semi-axes is given by $1/(1+a)^{2}$.

\begin{figure}
 \centering \scalebox{0.4}{\epsfig{file=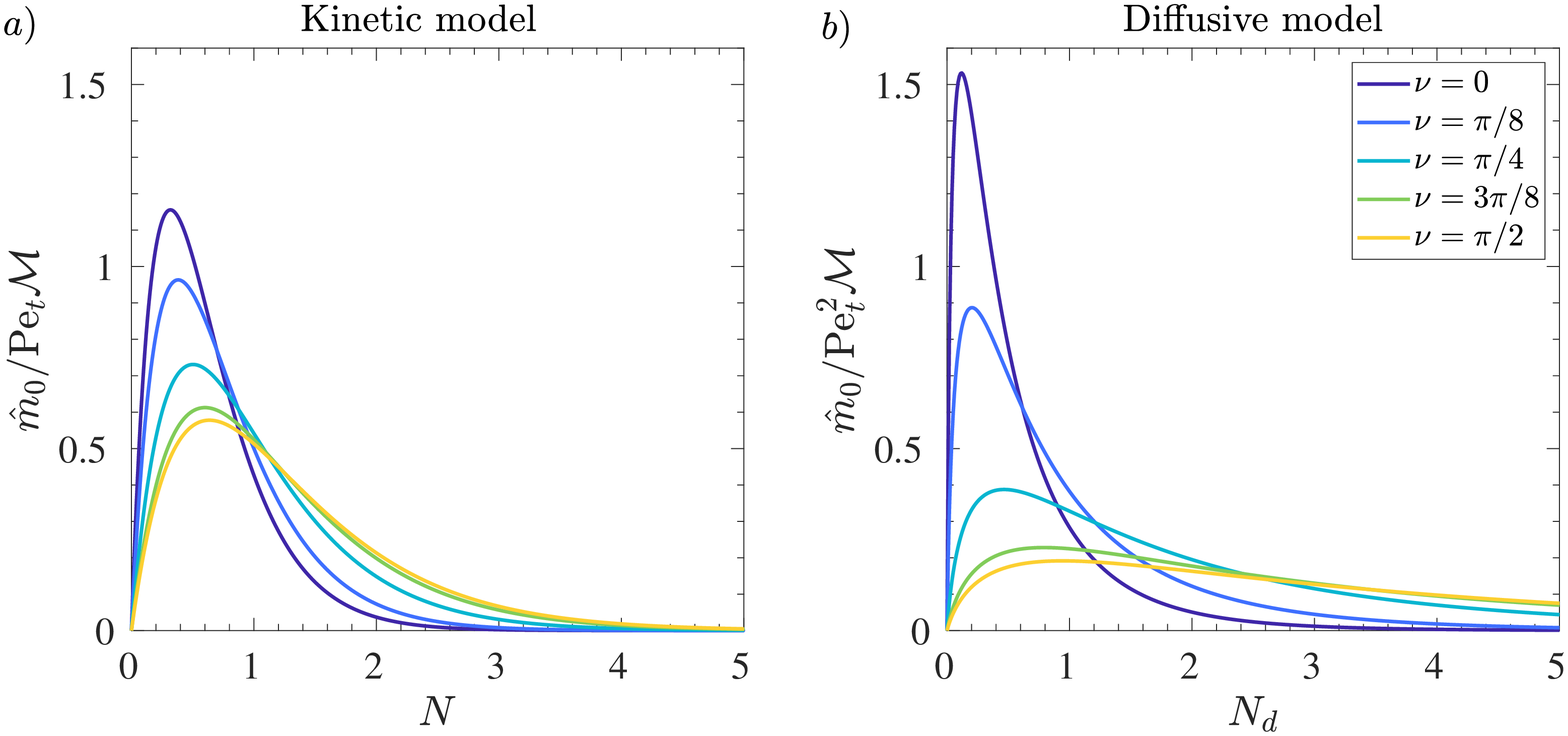}}
 \caption{Similarity profiles of the nascent coffee ring for an elliptical drop with $a = 1$ ($e(1) = 0.866$) for $a)$ the kinetic evaporative regime as given by (\ref{eqn:SimilarityEllipse}) and $b)$ the diffusive evaporative regime as given by (\ref{eqn:SimilarityEllipse_Diffusive}).} 
 \label{fig:SimilarityPlots}
 \end{figure}

All of these features can be seen by plotting the similarity profile (\ref{eqn:SimilarityEllipse}) for a droplet with $a = 1$ for different values of $\nu$, which is shown  in figure \ref{fig:SimilarityPlots}a: as $\nu$ increases, the mass profiles get progressively shallow and broader, with  the location of the peak progressively moving away from the contact line at $n = 0$. The accumulated mass flux $\mathcal{M}_{0}(\nu,0,t)$ simply acts to accentuate the coffee-ring height, with the peak location and the full-width at half-maximum remaining unchanged. 

A previous study by \citet{FreedBrown2015} demonstrated numerically that the mass flux of solute is stronger along the semi-major axis of a uniformly evaporating elliptical droplet. Here we have expanded upon this study, clearly illustrating that this increased mass flux combines with the effect of the local droplet profile leading to an enhanced coffee-ring along the more highly-curved parts of the contact line. Moreover, this is for a constant evaporative flux, showing that this asymmetry in the solute distribution can be driven by geometry-induced flow alone, which may have useful applications in evaporative-driven patterning processes \cite[][]{Harris2007}.

\begin{figure}
\begin{subfigure}
\centering \scalebox{0.32}{\epsfig{file=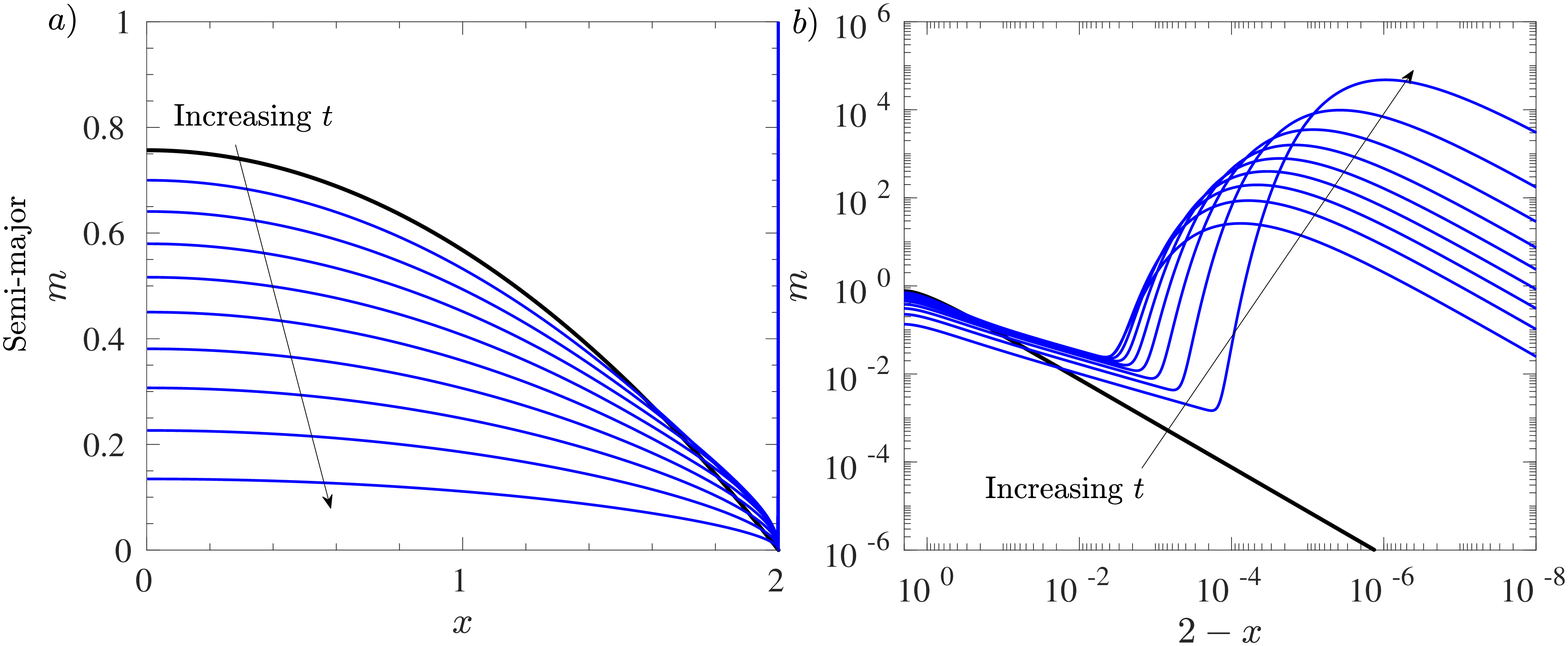}}
\captionsetup{labelformat=empty}
\end{subfigure}
\begin{subfigure}
\centering \scalebox{0.32}{\epsfig{file=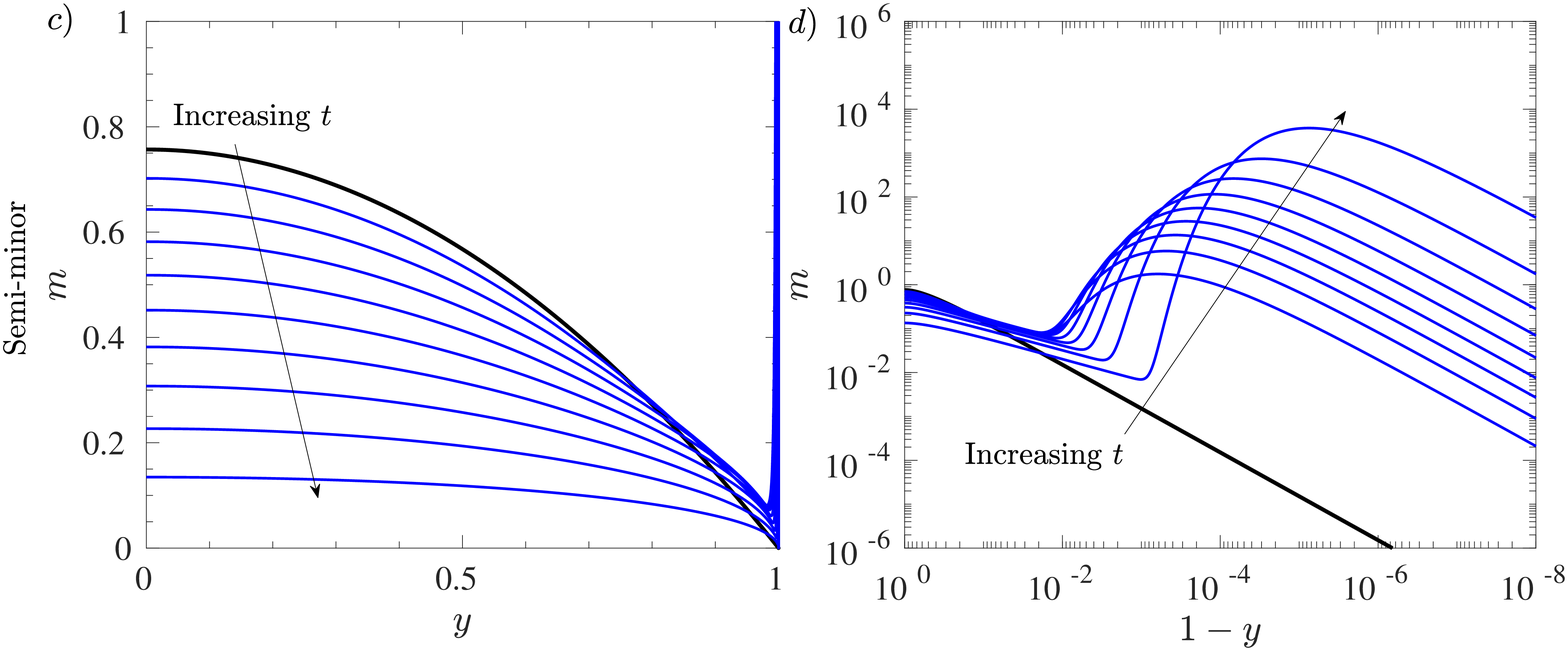}}
\end{subfigure}
\caption{Profiles of the solute mass $m = h\phi_{\mathrm{comp}}$ along the semi-major ($a, b$) and semi-minor ($c, d$) axes of an elliptical droplet with $a = 1$ ($e(1) = 0.866$) evaporating under a diffusive flux for $\Pe \approx 35$. In each figure, the bold, black curve represents the initial mass profile, while plots at time intervals of 0.1 up to $t = 0.9$ are shown as solid, blue curves.}
\label{fig:ProfileComparisons_Diffusive} 
\end{figure}

\subsubsection{Diffusive evaporation}

If one also allows the evaporative flux to vary as a result of the droplet geometry, the asymmetries in the nascent coffee ring become more exaggerated. To show this, we plot in figure \ref{fig:ProfileComparisons_Diffusive} mass profiles along the minor and major semi-axes for an ellipse with $a = 1$ ($e(1) = 0.866$) evaporating under a diffusive flux (the droplet again has the same initial volume and perimeter as that in figures \ref{fig:Validation} and \ref{fig:ProfileComparisons}). The coffee ring effect is significantly enhanced by the contact line geometry: the peak of the ring at $90\%$ of the drying time is $\approx 13$ times larger along the semi-major axis than the semi-minor axis. Moreover, the peak is also $\approx 10\%$ times larger than that in the equivalent axisymmetric diffusive problem. 

We illustrate how the properties of the nascent coffee ring vary with ellipse eccentricity in figure \ref{fig:Fixed_V_and_P_kinetic}c,d, where we show how the ring peak and its distance from the contact line change with $e(a)$ for a fixed initial droplet volume and perimeter. The results are displayed at $90\%$ of the drying time, with the P\'{e}clet number of the equivalent axisymmetric drop taken to be $\Pe_{0} = 200$ and the corresponding parameters for the droplets with elliptical footprints evaluated from (\ref{eqn:R_and_d_ratios}).

The results behave in a qualitatively similar manner to the kinetic regime. In particular, the peak height along semi-major axis initially increases as we increase the eccentricity, before reaching a maximum. For the diffusive case, the maximum is at $a \approx 0.5$ ($e(a) \approx 0.745)$, with the peak approximately $20\%$ larger than the equivalent axisymmetric droplet. As the eccentricity increases further, the peak height begins to decrease again, eventually decreasing below the axisymmetric peak. It is worth noting, however, that the width of the coffee ring is larger than in the equivalent axisymmetric droplet. Hence, for larger eccentricities, we have a shallower but wider coffee ring, as necessitated by the overall increased accumulated mass flux into the contact line (cf. figure \ref{fig:MassFluxes}).

For a diffusively evaporating droplet, the similarity profile can be found from (\ref{eqn:Similarity_Diff}) to be given by
\begin{linenomath}
 \begin{eqnarray}
  \frac{\hat{m}_{0}}{\Pe_{t}^{2}\mathcal{M}_{0}(\nu,0,t)} & = & \frac{\pi^{4}(1+a)^{10}}{48K[e(a)]^{4}}\frac{N_{d}}{(1+(2a+a^{2})\sin^{2}\nu)^{3}} \times \nonumber \\
  & & \mbox{exp}\left(\frac{-\pi(1+a)^{5/2}\sqrt{N_{d}}}{K(e(a))[1+(2a+a^{2})\sin^{2}\nu]^{3/4}}\right),
  \label{eqn:SimilarityEllipse_Diffusive}
 \end{eqnarray}
\end{linenomath}
where $N_{d} = \Pe_{t}^{2}n$. The coffee-ring peak and its location are given by
\begin{linenomath}
 \begin{alignat}{2}
  \frac{m_{\mathrm{max}}}{\Pe_{t}^{2}\mathcal{M}_{0}(\nu,0,t)} & \, = & \,  \frac{\pi^{2}(1+a)^{5}}{6\mbox{e}^{2}K[e(a)]^{2}}\frac{1}{[1+(2a+a^{2})\sin^{2}\nu]^{3/2}},\label{eqn:Peak_Height_Diffusive} \\ 
  \Pe_{t}^{2}n_{\mathrm{max}} & \, = & \, \frac{8K[e(a)]^{2}}{\pi^{2}(1+a)^{5}}\left[1+(2a+a^{2})\sin^{2}\nu\right]^{3/2}.\label{eqn:Peak_Location_Diffusive}
 \end{alignat}
\end{linenomath}

In the diffusive regime, even when we discount the accumulated mass flux into the contact line, there is a more significant strengthening of the coffee-ring effect compared to the kinetic evaporative model. Along the semi-major axis of the ellipse, the peak height is $(1+a)^{3}$ larger, $1/(1+a)^{3}$ closer to the contact line and $1/(1+a)^{3}$ thinner than along the semi-minor axis. This can clearly be seen in figure \ref{fig:SimilarityPlots}b, where we plot the similarity profile (\ref{eqn:SimilarityEllipse_Diffusive}) for different values of $\nu$. Generally speaking, even though there are variations with the angle-like variable $\nu$, the coffee ring in the kinetic regime is much more uniform than that in the diffusive regime.

\subsection{Summary}

Our findings corroborate those of \citet{Saenz2017}, who consider experiments and simulations of different-shape droplets evaporating under the diffusive evaporative model, demonstrating that there is a more pronounced coffee ring near the most highly-curved parts of the droplet contact line. While \citet{Saenz2017} attribute this to the asymmetry in the evaporative flux (specifically that it is stronger along these parts of the boundary), we have demonstrated that this is not the only factor: indeed the similarity profile (\ref{eqn:SimilarityEllipse_Diffusive}) shows that there is an enhanced coffee ring effect along the semi-major axis of an ellipse purely due to the flow asymmetry induced by the droplet geometry. Thus in the diffusive regime, it is a combined effect of the droplet geometry, the increased mass flux and the increased evaporative flux that contributes to the change in the coffee ring structure with contact line curvature.

\subsection{Limitations of the dilute regime}
\label{sec:Breakdown}

All of the above results hold under the assumption that the solute remains dilute as it evaporates. However, as solute is carried to the contact line, the concentration increases there. As a result, several effects that we have neglected, most notably concentration-dependent diffusivity and suspension viscosity, become relevant locally. Eventually, the solute may jam, leading to an effective moving of the fluid boundary inwards from the initial pinned edge. While we do not seek to investigate these effects in the current analysis, it is clearly important to understand the limitations of the dilute assumption. Moreover, even if the window over which the dilute model is valid is quite short, it is necessarily the first stage of the coffee ring formation, so the analysis presented here represents the early-time solution that will apply before these effects become relevant.

\subsubsection{Kinetic evaporation}

Let us suppose that the critical solute concentration at which finite concentration effects are important is given by $\phi_{c}^{*}$. This value will depend upon the solute under consideration. For the purposes of illustration, we shall assume that $\phi_{c}^{*} = 0.1$, while we shall take indicative values of the initial concentration from \citet{Deegan2000}, with $\phi_{\mathrm{init}}^{*}$ ranging from $10^{-6}-10^{-2}$.

Now, according to our asymptotic analysis, the maximum value of the solute concentration, $\phi_{\mathrm{max}}(t)$ occurs at the contact line. Thus, we can evaluate the composite profile (\ref{eqn:SoluteCompositeGeneral}) on $\partial\Omega$ to find that
\begin{linenomath}
 \begin{equation}
  \phi_{\mathrm{max}}(\nu,t) = \Pe^{2}\frac{\mathcal{M}_{0}(\nu,0,t)}{\theta_{c}(\nu,t)^{3}} + \Pe\frac{\gamma B(\nu,t)}{\theta_{c}(\nu,t)}+ \lim_{\substack{(x,y)\rightarrow\partial\Omega \\ n\rightarrow 0}} \left(\phi_{0}(x,y,t) - \frac{B(\nu,t)}{n}\right)
  \label{eqn:phimax}
 \end{equation}
\end{linenomath}
where $\gamma$ is the Euler-Mascheroni constant.

\begin{figure}
 \centering \scalebox{0.4}{\epsfig{file=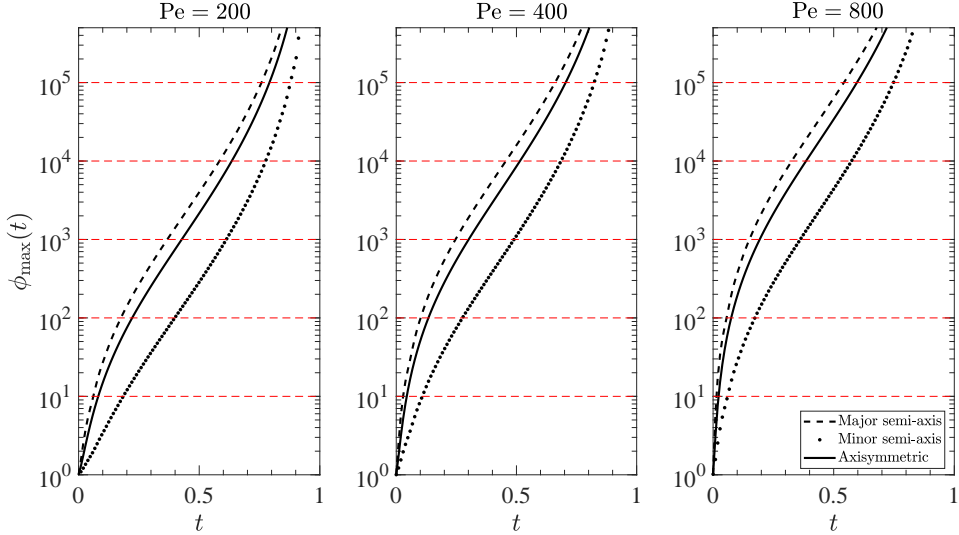}}
 \caption{The maximum solute concentration, $\phi_{\mathrm{max}}(\nu,t)$, for an elliptical droplet with $a = 1$ ($e(a) = 0.866$) evaporating under a constant evaporative flux as given by (\ref{eqn:phimax}). In each plot, the dashed curve displays the maximum concentration on the semi-major axis, $\phi_{\mathrm{max}}(0,t)$, the dotted curve displays the concentration on the semi-minor axis, $\phi_{\mathrm{max}}(\pi/2,t)$ and the solid curve shows the equivalent results for an axisymmetric droplet of the same volume and perimeter. The results are shown for $\Pe = 200, 400, 800$, where the P\'{e}clet number is defined with respect to the axisymmetric droplet. In each figure, the horizontal dashed red lines denote particular values of the critical concentration $\phi_{c}$. For each curve, the dilute regime lies to the \textit{left} of the intersection with the red lines.} 
 \label{fig:Jamming}
 \end{figure}

Note that $\phi_{\mathrm{max}}$ depends on the elliptical polar angle; our previous analysis has indicated that the maximum is higher along the semi-major axis of the ellipse than the minor and we expect this to directly translate into a reduced range of validity of the dilute model. To investigate this, in figure \ref{fig:Jamming} we plot $\phi_{\mathrm{max}}(\nu,t)$ as a function of time for an elliptical droplet with $a = 1$ and for different P\'{e}clet numbers along each semi-axis . In each figure, the value of $\phi_{\mathrm{max}}(\nu,t)$ along the semi-major axis is given by the dashed curve, while its value along the semi-minor axis is given by the dotted curve. For reference, the equivalent maximum concentration for an axisymmetric droplet of the same volume and perimeter is shown in each figure by the solid curve. The figures show the results for, from left to right, $\Pe = 200, 400, 800$, which are defined \textit{with respect to the axisymmetric droplet}, and the corresponding P\'{e}clet number for the elliptical case can be found using (\ref{eqn:R_and_d_ratios}). To help interpret the results, we have also included reference values of $\phi_{c} = \phi_{c}^{*}/\phi_{\mathrm{init}}^{*}$ in each figure as the dashed red lines. 

As is clearly seen in the figures, our intuition is correct: for a given value of $\phi_{c}$, the solute concentration along the semi-major axis reaches the critical value much earlier than along the semi-minor axis. Moreover, compared to an axisymmetric droplet of the same volume and contact line length, the critical solute concentration is reached sooner along the semi-major axis and later on the semi-minor axis. For a fixed $\Pe$, the time window over which the dilute regime remains valid increases as $\phi_{c}$ increases, while for a fixed $\phi_{c}$, the time window decreases as $\Pe$ increases. 

To take a concrete example, let us consider $\phi_{c} = 10^{5}$, which corresponds to a solute that is initially extremely dilute compared to the critical concentration. For $\Pe = 200$, the dilute regime is valid for $\approx80\%$ of the drying time for an axisymmetric drop, while for an elliptical drop, the dilute regime breaks down after $\approx 76\%$ of the drying time along the semi-major axis and after $\approx 87\%$ of the drying time along the semi-minor axis. On the other hand, for $\Pe = 800$, the dilute regime is valid for $\approx60\%$ of the drying time for an axisymmetric drop, while it breaks down after $\approx55\%$ of the drying time along the semi-major axis and after $\approx75\%$ of the drying time along the semi-minor axis for the equivalent elliptical drop. 

It is of note that in each of these cases, the dilute regime takes up a large percentage of the total drying time, indicating that the asymptotic analysis we have presented here gives a very good account of the nascent coffee ring formation and, in particular, allows us to predict the solute mass distribution within the droplet when finite concentration effects start to become relevant. Moreover, the solute mass profiles presented in, for example, \textsection \ref{sec:MassProfiles}, would be apt for comparison to experimental measurements of transient coffee ring profiles for an elliptical droplet evaporating under a kinetic evaporative flux.

However, it is worth noting that the time windows over which the dilute regime is valid do significantly reduce as $\phi_{c}$ decreases. For $\Pe = 200$ and $\phi_{c} = 10$, breakdown occurs after $\approx8\%$ of the drying time for the axisymmetric droplet, $\approx5\%$ of the drying time along the semi-major axis of the equivalent elliptical droplet and $\approx18\%$ of the drying time along the semi-minor axis. Nevertheless, our analysis is appropriate for the time window before breakdown, and provides the initial conditions for the regime in which finite concentration effects are relevant, so is likely to be an important consideration in understanding the characteristics of the final coffee ring.

We should also note that it may be that finite concentration effects are extremely localized in the model, so that, for example, even in problems where they are relevant close to the highly-curved parts of the contact line, the dilute model may still give an excellent description of the coffee ring dynamics for other parts of the boundary. This seems particularly reasonable given that the solute simply follows the streamlines in the droplet bulk (where the dilute regime is still valid), and the streamlines are independent of $t$. However, these comments do depend on the type of model chosen to incorporate finite concentration effects and whether such a model causes non-local changes to the liquid flow. We do not seek to address these questions any further here.

\subsubsection{Diffusive evaporation}

\begin{figure}
 \centering \scalebox{0.4}{\epsfig{file=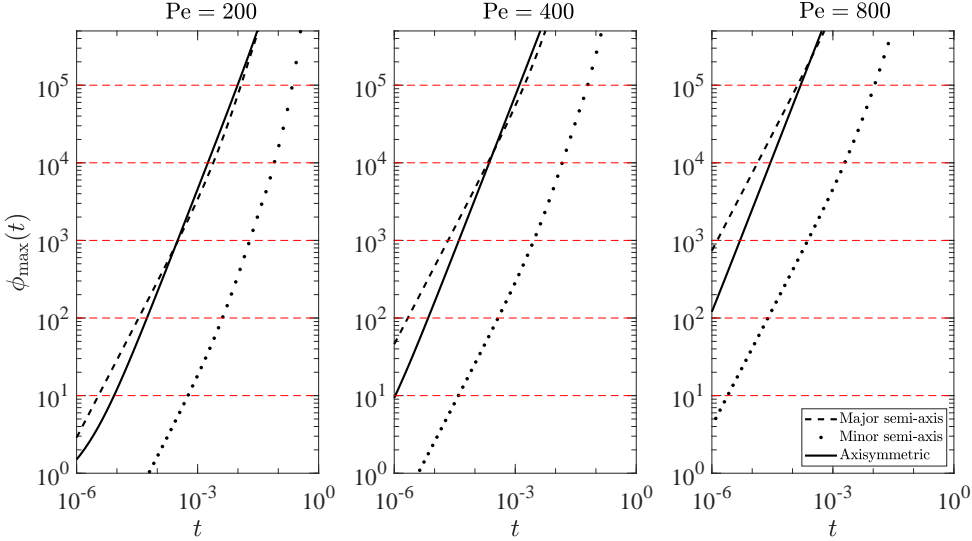}}
 \caption{The maximum solute concentration, $\phi_{\mathrm{max}}(\nu,t)$, for an elliptical droplet with $a = 1$ ($e(a) = 0.866$) evaporating under a diffusive flux as given by (\ref{eqn:phimax_Diffusive}). In each plot, the dashed curve displays the maximum concentration on the semi-major axis, $\phi_{\mathrm{max}}(0,t)$, the dotted curve displays the concentration on the semi-minor axis, $\phi_{\mathrm{max}}(\pi/2,t)$ and the solid curve shows the equivalent results for an axisymmetric droplet of the same volume and perimeter. The results are shown for $\Pe = 200, 400, 800$, where the P\'{e}clet number is defined with respect to the axisymmetric droplet. In each figure, the horizontal dashed red lines denote particular values of the critical concentration $\phi_{c}$.} 
 \label{fig:Jamming_Diffusive}
 \end{figure}

By the composite solution (\ref{eqn:SoluteCompositeDiffusive}), the maximum concentration in the diffusive evaporative regime is given by
\begin{linenomath}
 \begin{alignat}{2}
  \phi_{\mathrm{max}}(\nu,t) & \; = && \; \Pe^{4}\frac{64\chi(\nu)^{4}\mathcal{M}_{0}(\nu,0,t)}{3\theta_{c}(\nu,t)^{5}} + \Pe\frac{4\gamma \chi(\nu)B_{d}(\nu,t)}{\theta_{c}(\nu,t)} + \nonumber \\
  & \; && \; \lim_{\substack{(x,y)\rightarrow\partial\Omega \\ n\rightarrow 0}} \left(\phi_{0}(x,y,t) - \frac{B_{d}(\nu,t)}{\sqrt{n}}\right).
  \label{eqn:phimax_Diffusive}
 \end{alignat}
\end{linenomath}
It is immediately apparent that this is two orders of magnitude larger than (\ref{eqn:phimax}) and, accordingly, the time window over which the dilute regime is valid for this mode of evaporation is much smaller, as can be seen in figure \ref{fig:Jamming_Diffusive}. For $\Pe = 200$ and $\phi_{c} = 10^{5}$, we see that breakdown occurs for an axisymmetric droplet at just $\approx1\%$ of the drying time with this value rapidly decreasing as $\phi_{c}$ decreases or $\Pe$ increases. Breakdown along the semi-major axis of the equivalent ellipse occurs at a similar time, while there is an appreciable increase in the validity of the dilute model along the semi-minor axis to $\approx 23\%$ of the drying time. Again, this sharply tails off as $\phi_{c}$ decreases. Hence, if finite concentration effects can be treated locally, we may see a sizeable timeframe over which the dilute model presented here gives an accurate description of the nascent coffee ring along parts of the contact line with smaller curvature. Overall, however, it is clear that we need to consider these effects much sooner in the diffusive evaporative flux regime.

It is noticeable that there is marginal difference between the breakdown time along the semi-major axis and the equivalent axisymmetric droplet and indeed some cases where it appears to be reduced along the semi-major axis compared to the axisymmetric case. Given the extremely small timeframes under consideration, this is likely a combined effect of the numerical sensitivities in evaluating $\mathcal{M}_{0}(\nu,0,t)$ and $B_{d}(\nu,t)$ and the fact that the equivalent P\'{e}clet numbers are significantly smaller for an elliptical droplet (cf. Equation (\ref{eqn:R_and_d_ratios})).


\section{Summary and discussion}
\label{sec:Summary}

In this paper, we have presented a systematic asymptotic analysis of the solute profile as a thin, surface tension-dominated droplet of arbitrary contact set evaporates in the limit of large solutal P\'{e}clet-number, $\Pe\gg1$. Throughout, we have assumed that the droplet contact line remains pinned as the droplet evaporates. To illustrate the mathematical methodology, we focussed on two particular evaporation models: a simplified kinetic evaporation model in which the flux is uniform across the droplet free surface, and a diffusive evaporation model in which the flux is singular at the contact line. In the former case, we were able to isolate the effect of the droplet geometry alone on the nascent coffee ring characteristics, while for the latter regime, we were able to investigate the combined effects of the droplet geometry and an inhomogeneous evaporative flux.

Our analysis builds upon our previous work \cite[][]{Moore2021}, which revealed that it is the competing effects of solute diffusion and advection local to the contact line that drives the formation of the characteristic coffee-ring profile in the early stages of evaporation. In a more general geometry, the analysis is significantly more challenging,  but we were able to make asymptotic progress by utilizing a local orthogonal coordinate system $(s,n)$ that is embedded in the droplet contact line. This allows us to solve the leading-order local solute transport problem explicitly. To match with the advection-dominated region of the droplet, we exploited a formulation in terms of an integrated mass variable, which revealed that the local coffee ring profile is approximately a similarity profile $\hat{m}_{0}$ that is given by 
\begin{linenomath}
 \begin{equation}
  \frac{\hat{m}_{0}(s,N,t)}{\Pe_{t}\mathcal{M}_{0}(s,0^{+},t)} = f\left(N;2,\frac{1}{\psi(s)}\right), \; N = \Pe_{t}n
  \label{eqn:SimilarityForm_Conc}
 \end{equation}
\end{linenomath}
for a kinetic evaporative flux, and by
\begin{linenomath}
 \begin{equation}
  \frac{\hat{m}_{0}(s,N_{d},t)}{\Pe_{t}^{2}\mathcal{M}_{0}(s,0^{+},t)} = \frac{2\chi(s)}{3\psi(s)}f\left(\sqrt{N_{d}},3,\frac{4\chi(s)}{\psi(s)}\right), \quad N_{d} = \Pe_{t}^{2}n
  \label{eqn:Similarity_Diff_Conc}
 \end{equation}
\end{linenomath}
for a diffusive evaporative flux. In equations (\ref{eqn:SimilarityForm_Conc})--(\ref{eqn:Similarity_Diff_Conc}), $\mathcal{M}_{0}(s,0^{+},t)$ is the mass accumulated at the contact line, $\theta_{c}(s,t) =(1-t)\psi(s)$ is the local contact angle of the droplet, $\chi(s) \sim E(s,n)n^{1/2}$ is the strength of the singularity in the local evaporative flux in the diffusive regime, $f(x;k,l) = l^{k}x^{k-1}e^{-lx}/\Gamma(k)$ is the probability density function of a gamma distribution, $\Pe_{t} = \Pe/(1-t^{*}/t_{f}^{*})$ is the modified P\'{e}clet number and $t_{f}^{*}$ is the dimensional dryout time of the drop. Characteristics of the nascent coffee ring such as the ring height and width can then readily be found from these similarity profiles.

Equations (\ref{eqn:SimilarityForm_Conc}) and (\ref{eqn:Similarity_Diff_Conc}) display the characteristic narrow, peaked profile of the nascent coffee ring and it is notable that this profile is dependent on the location on the contact line through the coordinate $s$. Hence, asymmetry in the droplet profile, the evaporative flux and the rate at which solute mass is transported to the contact line may all contribute to variation in the nascent coffee ring profile. 

After validating our asymptotic analysis in the axisymmetric regime by comparing to numerical simulations, we moved on to consider the example of a thin droplet with an elliptical contact set. For both evaporative models, the flow in the ellipse is stronger towards the semi-major axis where the contact line curvature is higher, and this effect is accentuated as the eccentricity of the ellipse increases. The increased velocity coupled with the geometry of the local free surface profile leads to an increase in the accumulated mass flux into the more highly-curved part of the elliptical boundary. We showed that, for droplets of identical volume and contact line perimeter, increasing the ellipse eccentricity increases (respectively, decreases) the accumulated mass flux into the boundary along the major (minor) semi-axis. This effect was more pronounced (although of a similar order of magnitude) for a diffusive evaporative flux. The increased accumulated mass flux contributes to a strengthened (respectively, weakened) coffee ring along the major (minor) semi-axis. 

It is notable that this effect is exhibited by both the kinetic and diffusive evaporative models. In particular, while it certainly contributes if present, asymmetry in the evaporative flux is not necessary to observe a variation in the coffee ring effect. This qualifies the conclusion of \citet{Saenz2017} that attributes coffee ring asymmetry due to an inhomogeneous evaporative flux.

For both evaporative models, the decreased mass flux into the contact line along the semi-minor axis manifests itself as a shallower, wider coffee ring than an equivalent axisymmetric droplet of the same initial volume and perimeter. The behaviour along the semi-major axis is richer. As the eccentricity of the droplet contact set initially increases, for both evaporative models the coffee ring becomes higher and narrower. The height reaches a maximum before falling in both evaporation models, with the effect starker for diffusive evaporation. However, to compensate for the increased mass flux, the ring then starts to broaden as compared to the axisymmetric droplet. As the eccentricity of the ellipse approaches unity, in the kinetic model, we found that the coffee ring height was comparable to the axisymmetric droplet, but that the ring was thicker, while for the diffusive model, we found that the height was in fact lower than the axisymmetric droplet, but the ring was much thicker.

We concluded by using our asymptotic results to investigate when the dilute regime breaks down and finite concentration effects are likely to become relevant close to the contact line, where the solute concentration is maximal. As may be expected, the enhanced flow and coffee-ring effect along the semi-major axis reduces the time window over which the dilute model is valid as compared to the equivalent axisymmetric droplet. However, this effect is very much localized: indeed, along the semi-minor axis, the time window is correspondingly lengthened. Clearly finite concentration effects may be present in some parts of the droplet for a significantly longer period than others. It is notable that the dilute regime is valid for significantly longer for a kinetic evaporative flux compared to a diffusive flux: this is due to the significantly enhanced coffee-ring effect in the latter regime. This longer time period of validity coupled with the variable coffee ring effect along different parts of the contact line suggest that the kinetic evaporative model may be ripe for exploitation in engineering applications in which dynamically controlling the deposit shape is important, for example in colloidal patterning \cite[][]{Choi2010} and in printing conductors \cite[][]{Layani2009}.

Even in situations where the time window of applicability is relatively small, the dilute model necessarily applies in the early stages of coffee ring formation. Hence the analysis derived here provides a description of the flow profile and solute distribution before finite concentration effects are introduced. There are a number of different avenues that could be pursued to model such effects, whether through a simple jamming model \cite[such as that in][]{Popov2005}, accounting for the increasing concentration through suspension-dependent viscosity and diffusivity \cite[as in ][]{Kaplan2015} or through more complicated two-phase suspension models \cite[see, for example,][]{Guazzelli2018}. These are all interesting avenues for future studies.

\tbf{Declaration of Interests.} The authors report no conflict of interests.


\appendix
\section{Numerical methods}
\label{appendix:Numerics}

In this appendix, we describe the numerical approaches necessary to solve for the liquid flow and the solute concentration for thin droplets with an elliptical contact set. We shall present the methodology for the kinetic evaporative model in which $E = 1$, but the methodology extends readily to the diffusive regime, as discussed presently.

\subsection{Solution of the pressure problem for a droplet with an elliptical contact set}

To find the liquid velocity, we must find the pressure perturbation $P(x,y)$ that satisfies the Neumann problem (\ref{eqn:PressurePert}). Let us first define
\begin{linenomath}
 \begin{equation}
  P = 8\left(1+\frac{1}{(1+a)^{2}}\right)^{3}\left(\mathcal{P} + \mathcal{P}_{s}\right), \; \mathcal{P}_{s} = -\frac{3}{\mathcal{H}+|\nabla\mathcal{H}|^{2}}\frac{1}{\mathcal{H}}, \; \mathcal{H} = 1 - \left(\frac{x}{1+a}\right)^{2} - y^{2},
 \end{equation}
\end{linenomath}
which reduces (\ref{eqn:PressurePert}) to
\begin{linenomath}
 \begin{alignat}{2}
  \nabla\cdot\left(-\frac{\mathcal{H}^{3}}{3}\nabla\mathcal{P}\right) & \, = && \, 2\mathcal{H} - \left(3+\frac{2}{(1+a)^{2}}\right)\frac{\mathcal{H}}{\mathcal{H}+|\nabla\mathcal{H}|^{2}}  \nonumber \\
  &\, && \,- \mathcal{H}\nabla\mathcal{H}\cdot\nabla\left(\frac{1}{\mathcal{H}+|\nabla\mathcal{H}|^{2}}\right)-\mathcal{H}^{2}\nabla^{2}\left(\frac{1}{\mathcal{H}+|\nabla\mathcal{H}|^{2}}\right)
 \label{eqn:ScaledPressurePert1}
 \end{alignat}
\end{linenomath}
in $\Omega$, such that
\begin{linenomath}
 \begin{equation}
  -\frac{\mathcal{H}^{3}}{3}\nabla\mathcal{P}\cdot\mbf{n} = 0 \; \mbox{on} \; \partial\Omega. \label{eqn:ScaledPressurePert2}
 \end{equation}
\end{linenomath}
We have chosen the form of $\mathcal{P}_{s}$ to leave the boundary condition (\ref{eqn:ScaledPressurePert2}) unchanged while also subtracting out the $O(1/n)$ singularity in $P$ at the contact line (which improves convergence of the numerical solution).

\begin{figure}
\centering \scalebox{0.5}{\epsfig{file=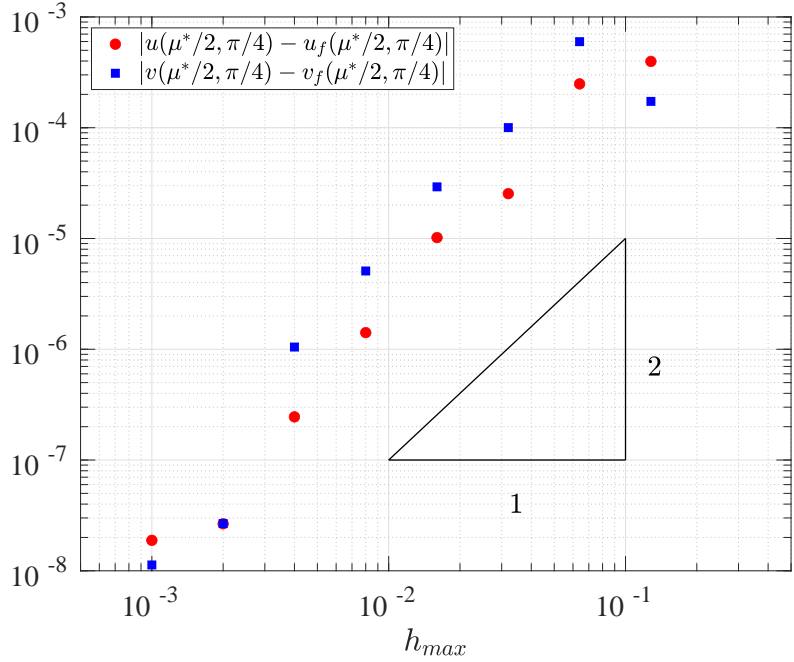}}
\caption{Doubly logarithmic plot of the relative error of the computed velocity components at the centre of the rectangle for various mesh sizes compared to the solution on an extremely fine mesh ($h_{max} = 8\times10^{-4}$).}
\label{fig:Convergence1}
\end{figure}

Using symmetry, we solve the problem for $x,y >0$ and, to simplify the domain, we solve in the planar elliptical coordinate system $(\mu,\nu)$, which is defined in (\ref{eqn:EllipticalPolars}). We solve (\ref{eqn:ScaledPressurePert1})--(\ref{eqn:ScaledPressurePert2}) using MATLAB's in-built finite element code in the PDE Toolbox. To show that the code converges, we consider the particular case when $a = 1$ and vary the value of $h_{\mathrm{max}}$, the maximum allowed size of an element in the simulation. In figure \ref{fig:Convergence1}, we display the absolute error in each velocity component at the centre of the rectangle for different values of $h_{\mathrm{max}}$ and the solution calculated on a very fine grid (for which $h_{max} = 8\times10^{-4}$). We see that the error decreases proportionally to $h_{\mathrm{max}}^{2}$ as we refine the grid, as anticipated. 

\subsection{Solution of the leading-order-outer solute problem}

To solve for the leading-order-outer solute mass as given by (\ref{eqn:OuterSolution}), we need to solve for the Jacobian $J$ using Euler's identity, (\ref{eqn:EulersIdentity}). We note that by setting
\begin{linenomath}\begin{equation}
 \tau = -\log\left(1-t\right), \; \bar{\mbf{u}} = \frac{\bar{\mbf{U}}}{1-t}
\end{equation}\end{linenomath}
where $\bar{\mbf{U}}$ is independent of $t$ and is given by the solution of (\ref{eqn:PressurePert}), we reduce the problem to solving
\begin{linenomath}
\begin{equation}
 \frac{\mbox{D}}{\mbox{D}\tau}\left(\log{J}\right) = t_{f}\nabla\cdot\bar{\mbf{U}} \quad \mbox{for} \; \tau > 0, 
 \label{eqn:ScaledEulers}
\end{equation}
\end{linenomath}
subject to $\log{J} = 0$ at $\tau = 0$.

Our methodology for solving this problem numerically is as follows. Firstly, we pick a location $\mbf{x}^{\dag}$ on the contact line. Then, following \citet{FreedBrown2015}, we can find the initial location of the point, $\mbf{x}_{0}$ say, that reaches $\mbf{x}^{\dag}$ at time $\tau = \tau^{\dag}$ by solving
\begin{linenomath}
\begin{equation}
 \frac{\mbox{D}\mbf{x}}{\mbox{D}\tau} = -t_{f}\bar{\mbf{U}} \quad \mbox{subject to} \; \mbf{x} = \mbf{x}^{\dag} \; \mbox{at} \;\tau = 0, \quad \mbf{x} = \mbf{x}_{0} \; \mbox{at} \; \tau  =\tau^{\dag}.
 \label{eqn:ScaledEulers2}
\end{equation}
\end{linenomath}
Once $\mbf{x}_{0}$ is found, we then find the value of the Jacobian at $\mbf{x}^{\dag}$ at $\tau = \tau^{\dag}$ by integrating (\ref{eqn:ScaledEulers}) along a streamline starting from $\mbf{x}_{0}$.

Since all of the equations (\ref{eqn:ScaledEulers}) and (\ref{eqn:ScaledEulers2}) are autonomous, they are relatively straightforward to solve using MATLAB's inbuilt $\tit{ode15s}$ solver. We do so with 2000 time stations clustered at times at which the velocity is largest (i.e. when we approach the contact line).

To find the coefficient of local concentration as given by (\ref{eqn:Phi_local}) in the kinetic regime and (\ref{eqn:DiffusiveOuterLocal}) in the diffusive regime, we choose a number of angular stations close to the contact line and then repeat the above procedure at dimensionless time intervals of $5$\% of the drying time up to $95$\% of the drying time. We then interpolate the data to obtain results at intermediate timesteps.


\bibliographystyle{jfm}

\bibliography{EllipticCoffeeRings}
 


\end{document}